\documentclass[prb,aps,tightenlines,twocolumn,
showpacs,floatfix]{revtex4}
\usepackage{graphicx}
\usepackage{amssymb}
\usepackage{amsmath}
\usepackage{bm}
\usepackage{colordvi}

\begin{document}
\title{Theory of excitons in cubic III-V semiconductor GaAs, InAs
and GaN  quantum dots: fine structure and spin relaxation}
\author{H. Tong}
\author{M. W. Wu}
\thanks{Author to whom correspondence should be addressed}
\email{mwwu@ustc.edu.cn.}
\affiliation{Hefei National Laboratory for Physical Sciences at
  Microscale and Department of Physics, University of Science and
  Technology of China, Hefei, Anhui, 230026, China}

\date{\today}
\begin{abstract}
Exciton fine structures in cubic III-V semiconductor GaAs, InAs and
GaN quantum dots
are investigated systematically and the exciton spin relaxation 
in GaN quantum dots is calculated by first setting up the effective
exciton Hamiltonian. The electron-hole
 exchange interaction Hamiltonian, which consists of the 
long- and short-range parts, is derived within the
effective-mass approximation by taking into account the conduction, 
 heavy- and light-hole bands, and especially the split-off band.
The scheme applied in this paper  allows the description of excitons 
in both the strong and weak confinement regimes.
The importance of treating
the direct electron-hole Coulomb interaction unperturbatively is
demonstrated. We show in our calculation that the light-hole and 
split-off bands are negligible
when considering the exciton fine structure, even for GaN quantum dots,
and the short-range exchange
interaction is irrelevant when considering the optically active doublet 
splitting. We point out that the long-range exchange interaction, which is
neglected in many previous works,   
contributes to the energy splitting between the bright and dark states,
together with the short-range exchange interaction.
 Strong dependence of the optically active doublet splitting
on the anisotropy of dot shape is reported. 
Large doublet splittings up to
600~$\mu$eV, and even up to several meV for  small dot 
size with large anisotropy, are shown in GaN quantum dots. 
The spin relaxation between the
lowest two optically active exciton states in GaN quantum dots is
calculated, showing a strong
dependence on the dot anisotropy. Long exciton spin relaxation time 
is reported in GaN quantum dots. 
These findings are in good agreement with the experimental results. 
\end{abstract}


\maketitle

\section{INTRODUCTION}
Semiconductor quantum dots (QDs) have attracted intense
interest due to the high 
potential to work as basic device units for photonics, spintronics, quantum
communication and computation.\cite{Barenco,Biolatti,Lovett,Boyle,Loss,Hanson,Klauser} One
of the applications is to use QDs as emitter of 
single photon\cite{Michler,Yuan,Shields,Santori2} and entangled photon
pairs\cite{Benson,Stev,Avron,Akopian,Greilich,Hafenbrak,Stevenson} via the combination of exciton
and biexciton. Explicitly, for a symmetric QD, the decay of the bright (optically active)
exciton states\cite{Ivchenko5} with total angular momentum projections $J_z=\pm
1$ can emit $\sigma^{\pm}$ circularly 
polarized photons, while that of  the biexciton states through two
intermediate degenerate bright exciton states can generate two entangled
photons.\cite{Benson,Mohan} The asymmetry of QDs lifts the degeneracy of the two 
bright exciton states and mixes them into two states that generate linearly
polarized photons with orthogonal polarizations during the 
decay.\cite{Akopian,Greilich,Santori,Gammon,Besombes} This is even crucial when considering the
biexciton decay as the splitting of the intermediate exciton states makes the two
channels distinguishable and hence destroys the entanglement.\cite{Avron,Santori,Ulrich,Stevenson}
Moreover, the dark (optically forbidden) exciton states which lie slightly below
the bright ones\cite{Ivchenko5} are considered promising candidates for spin qubits due to their
long lifetimes when confined in QDs.\cite{Imam,Johansen} They also play a key role
for the Bose-Einstein
condensation in semiconductors.\cite{Combescot}

The splitting between the bright and dark exciton states (denoted as the BD
exchange splitting hereafter)\cite{Nirmal,Bayer2} and that between the bright
exciton states (denoted as the doublet
  splitting hereafter)\cite{Seguin,Young,Stevenson2,Dovrat,Sugisaki,Seguin2}
 are mainly 
controlled by the electron-hole (e-h) exchange interaction together with the dot
size and asymmetry.\cite{Bayer1,Taka2,Zunger2,Kindel,Young,Besombes} So a 
detailed understanding of the e-h exchange interaction and the resulting exciton
fine structure in QDs is of fundamental importance for both
theoretical and application purposes. 

The e-h exchange interaction has been investigated ever since the 1960s. It can be
decomposed into long- and short-range parts in the real
space\cite{Knox,Bir1, Bir2} or the analytical and nonanalytical parts in the
${\bf k}$ space.\cite{Denisov,Ivchenko2} There is a close
correspondence between them which can be found in Ref.~\onlinecite{Cho1}, and sometimes
no difference is made between these two approaches.\cite{Ivchenko1}

Early investigations on the e-h exchange interaction mainly focus on the bulk
system.\cite{Heller,Denisov,Onodera,Abe,Elliott,Rossler,Bonneville,Suga,Andreani2}
It is well known that, when 
considering the exciton states in 
semiconductors by taking into account the conduction band $\Gamma_6^c$ and the 
valence band $\Gamma_8^v$, the short-range (SR) exchange interaction splits
the eightfold-degenerate exciton state into a triplet bright state and a quintuplet
dark state with the splitting energy between them, the so-called exchange
energy.\cite{Bonneville,Suga} The long-range (LR) exchange interaction further
splits the triplet into a 
longitudinal and two transverse modes, the energy difference of which is denoted as
the longitudinal-transverse splitting.\cite{Bonneville,Suga,Andreani2}
The e-h exchange interaction, as well as the direct Coulomb
interaction, is greatly enhanced by the quantum confinements in low-dimensional
semiconductor structures because of the increased spatial overlap between the
electron and hole wave functions. Reexamination of the e-h exchange interaction
in low-dimensional structures was intrigued in both 
experimental\cite{Bauer,Kesteren,Norris,Chamarro,Amand} and
theoretical\cite{Jorda,Ivchenko5,Chen,Andreani,Maialle,Taka1,Taka2,Efros,Horo,Kalt1} 
ways. Most of the
theoretical works were carried out within the framework of the 
envelope-function approximation together with the effective-mass
approximation.\cite{Chen,Andreani,Maialle,Taka1,Taka2,Efros,Horo,Kalt1} 
In general, due to the different effective masses of the heavy, light and
  split-off holes, the heavy-, light- and split-off-hole exciton states\cite{note0}
  are energetically split when a quantum confinement is applied.  For common
  cubic III-V semiconductors, the heavy-hole exciton which energetically
  lies the lowest is of most physical interest and is hence mostly
  investigated.\cite{Chen,Bauer,Andreani,Maialle,Kesteren,Kalt1,Kalt2,Kalt3,Ivchenko3,Kada}
  The heavy-hole exciton quartet, which is characterized by the total angular
  momentum projections $J_z=\pm 1$, $\pm2$, is split into bright and dark
  exciton states with $J_z=\pm 1$ and $J_z=\pm 2$, respectively, by the e-h
  exchange interaction and the bright exciton states $|\pm 1\rangle$ are further
  split under anisotropic confinement potential.
Chen {\em et al}.\cite{Chen} calculated the 
exchange energy and the BD exchange splitting in GaAs/${\rm
  Ga}_{1-x}{\rm Al}_x{\rm As}$ quantum wells (QWs) and evidenced 
the enhancement of the exchange effect with decreasing well width. Their
work was based on the approximation of
 decoupled heavy- and light-hole subbands.  
Andreani and Bassani investigated the ${\bf k}$ dependence of the e-h exchange
interaction in QW systems.\cite{Andreani} 
In Ref.~\onlinecite{Maialle} the exciton spin dynamics in 
GaAs QWs was studied with e-h exchange interaction as an effective spin-flip
mechanism, where the matrix elements of the LR exchange interaction were
calculated by  using the simple
heavy-hole exciton ground states while those of the SR
exchange interaction were obtained by taking into account the mixing of
heavy- and light-hole bands. Takagahara performed 
systematic studies of exciton states in QDs in Ref.~\onlinecite{Taka1}. The
exciton binding energy and the LR and SR exchange interactions were investigated
using wave functions
calculated by the variational method. The subband
mixing induced by the direct Coulomb
interaction was pointed out to be important. In a later
work for GaAs QDs,\cite{Taka2} Takagahara derived an eight-band exciton
Hamiltonian. The LR exchange interaction, which was
attributed to dipole-dipole interaction, was 
emphasized to be much more important than  the SR exchange interaction when
concerning the exciton doublet fine structure. The contradiction of the
  calculated scaling law of the doublet splitting energy to the theoretical
  prediction was reported (also in Ref.~\onlinecite{Kada}).
  Whereas in the latter works by Tsitsishvili {\em et
      al}.\cite{Kalt1,Kalt2} the exciton spin relaxation in 
single asymmetrical QD was studied by taking into account only the 
SR exchange interaction. Efros {\em et
  al}.\cite{Efros} and Horodysk\'{a} {\em et al}. in their very recent work\cite{Horo} investigated the
band-edge exciton states in spherical QDs by including only
the SR exchange interaction. 
But actually, as will be shown in this paper, the LR exchange interaction contributes
to the splitting between bright and dark exciton states even in isotropic QDs
where the doublet splitting energy vanishes. Its contribution to the 
BD exchange splitting  can be comparable with that from the SR 
exchange interaction.
Moreover, although the direct Coulomb interaction is believed to lead to
 poor convergence if treated perturbatively when the exciton system is in the
weak confinement regime, in a very recent work by Kadantsev and
  Hawrylak,\cite{Kada} the exciton fine structure in 
GaAs QDs was still studied by taking the
 direct Coulomb interaction as perturbation.

Thus, even though a lot of works have been done on the e-h exchange
interaction induced exciton properties, obvious confusions are still seen in the
literature, even for the most investigated In$_{1-x}$Ga$_x$As
nanostructures. The significance of valence-band coupling to the exciton fine
structure needs to be evaluated. The relative importance of the LR and SR exchange
interactions to the doublet splitting needs more elaboration, and that to the BD
exchange splitting needs to be clarified. Furthermore, the size scaling of the
e-h exchange interaction and the influence of the direct
Coulomb interaction on the exciton fine structure have to be examined and stressed.

Works discussed above mainly concern III-V semiconductor nanostructures based on
In$_{1-x}$Ga$_x$As the 
split-off band of which is far away from the heavy- and light-hole 
bands.\cite{Vurga} So the split-off band is always neglected when studying  
exciton fine structure. However, for cubic GaN, whose
spin-orbit splitting is small compared to the wide band
gap,\cite{Vurga} no theoretical work has been performed to investigate exciton
fine structure so far, nor does the explicit
expression of e-h exchange Hamiltonian with the effect of 
split-off band exists in the literature. It has been proved that in bulk GaN, 
the split-off band is important when considering the spin-orbit
coupling.\cite{Fu} Whether it is still the case when studying the e-h exchange
interaction in GaN QDs needs to be examined.

The exciton spin relaxation is another important subject that strongly affects
the quality of applications in information storage and processing based on
exciton states in QDs.\cite{Kalt1,Kolli,Smith,Astakhov}
Motivated by recent experimental study of exciton spin orientation in cubic
GaN/AlN QDs,\cite{Marie} we investigate the behavior of  the LR and SR
exchange interactions in cubic III-V semiconductor QDs, 
taking into account all heavy-hole, light-hole and split-off bands. The spin
relaxation between the lowest two bright exciton states in single
GaN QD is studied after that. 

In this paper, in order to explicitly include the effect of the split-off band on
the exciton fine structure in cubic GaN, GaAs and InAs QDs, the bulk e-h exchange 
Hamiltonian of both LR and SR parts is first derived in the $12\times 12$
matrix representation. The derivation
is carried out within the framework of the effective-mass approximation.\cite{Winkler} For
systems strongly confined in one direction, e.g., the QD system with small dot
height considered in this paper, we are 
able to apply the L\"owdin partitioning method\cite{Winkler, 
  Lowdin, Luttinger2} to approximately diagonalize the modified $6\times 6$
Luttinger Hamiltonian (specified in Sec.~II) for holes to obtain a 
new ``heavy-hole'' band, which is an admixture of the heavy-hole, light-hole and
split-off bands. In this way, $4\times 4$ matrix representations of the
exciton exchange Hamiltonian are constructed by taking the 
conduction band and the new ``heavy-hole'' band with the effect of
valence-band mixing included. We then apply the effective
Hamiltonian obtained to investigate the exciton fine structures in cubic III-V
semiconductor QDs. The doublet splitting energy and the BD exchange
splitting are calculated and the relative importance
of  the LR and SR exchange 
interactions as well as that of
 the heavy-hole, light-hole and split-off bands are discussed.  The
size scalings of the doublet  
splitting energy and the BD exchange splitting contributed by the LR and the SR
exchange interactions are analyzed and explained by the scaling rules
established.

Due to the fact that the values of the exciton Bohr radius in bulk
GaAs, InAs and GaN are 14.9~nm, 51.6~nm and 4.8~nm,\cite{Ekardt} respectively, 
which are comparable or even 
smaller than the average diameter of the QDs in this paper (given in
Sec.~III), chosen according to the 
experiments,\cite{Marie,Bayer1,Bayer2,Raymond} 
the direct Coulomb interaction is too large to be treated perturbatively. 
We solve the Schr\"odinger equation by taking the  
direct Coulomb interaction and the confinement in an equal footing. The
calculated exciton binding energies are found to be considerably large,
markedly enhanced by the confinement. This 
contradicts the results in the latest work by Kadantsev and
  Hawrylak\cite{Kada} and further demonstrates the
importance to treat the direct Coulomb interaction unperturbatively. 
The importance of
the direct Coulomb interaction to the exciton fine structure is demonstrated.
Finally, the exciton spin relaxation
rates in single GaN QDs are calculated and long spin
 relaxation time is obtained. Our results
are in agreement with experiments.\cite{Bayer1,Seguin,Marie}

This paper is organized as follows: In Sec.~II, we set up our model and lay out
the formalism. Matrix representations of LR and SR exchange interactions are
derived first in bulk and then in QW. Size-scaling rules are
established and exciton spin relaxation assisted by the acoustic phonons are
introduced after that. The numerical scheme is laid out at the end of this
  section. In Sec.~III, we show our numerical
results of exciton fine structures in GaAs, InAs and GaN QDs. Explicit
properties of the LR and SR exchange interactions contributing to the doublet
splitting energy and BD exchange splitting are discussed in detail. 
In Sec.~IV, the
exciton spin relaxation in single GaN QD is investigated.
 We conclude in Sec.~V.

\section{MODEL AND FORMALISM}
We formulate the theory of excitons in cubic III-V
semiconductors by taking into 
account the conduction band $\Gamma_6^c$ and the valence bands $\Gamma_8^v$ and
$\Gamma_7^v$. The effective representations of the
 exciton Hamiltonian for the LR
and SR exchange interactions are derived first in bulk and then in QW. 
The size-scaling rules of the e-h exchange interaction are established after
that. We then introduce the exciton spin relaxation due to the
electron/hole--acoustic-phonon scattering. The numerical scheme is laid out
at the end of this section.
\subsection{e-h exchange interaction}
\subsubsection{e-h exchange interaction in  bulk }
We start our investigation on the e-h exchange interaction from a
general description of direct Wannier-Mott excitons in bulk system within the
framework of effective-mass approximation.
The exciton wave function can be written in the form\cite{Bir1,Bir2}
\begin{eqnarray}
\nonumber
  \Psi({\bf r}_1,{\bf r}_2) = \sum_{mn}\big[F_{mn}({\bf r}_1,{\bf r}_2)\psi_{m{\bf
      k}_0}({\bf r}_1)\tilde{\psi}_{n{\bf k}_0}({\bf r}_2) \\
\mbox{}-F_{mn}({\bf r}_2,{\bf r}_1)\psi_{m{\bf
      k}_0}({\bf r}_2)\tilde{\psi}_{n{\bf k}_0}({\bf r}_1)\big],
\label{equ:Wavef1}
\end{eqnarray}
where $\psi_{m{\bf k}_0}({\bf r})$ is the conduction-band Bloch function and
$\tilde{\psi}_{n{\bf k}_0}({\bf r})$ is the Bloch function for the hole which
is the time reversal of the Bloch function of the missing
electron.\cite{Bir1,Bir2} As for cubic III-V semiconductors, we are 
interested in excitons at the band edge, i.e., the $\Gamma$ point with
${\bf k}_0=0$. $m$ ($n$) is the index for the electron 
(hole) band under consideration,
including the spin degree of freedom. $F_{mn}({\bf r}_1,{\bf r}_2)$ is the envelope
function with $F_{mn}({\bf r}_1,{\bf r}_2) = -F_{nm}({\bf r}_2,{\bf r}_1)$.  
This makes the exciton wave function antisymmetric.

The eigen equation for the envelope function $F_{mn}$ is given by\cite{Bir1,Bir2}
\begin{eqnarray}
  \nonumber
  && \sum_{mn} \int d{\bf r}_1d{\bf r}_2 H^{\rm eh}_{m^\prime n^\prime \atop mn}\left({\bf r}^\prime_1\ {\bf r}^\prime_2 \atop 
      {\bf r}_1\ {\bf r}_2\right) F_{mn}({\bf r}_1,{\bf r}_2)  \\
  &&\ \ \ \ \ \ \  = E F_{m^{\prime}n^{\prime}}({\bf r}^{\prime}_1,{\bf r}^{\prime}_2).
      \label{equ:SE1}
\end{eqnarray}
The explicit form of 
$H^{\rm eh}_{m^{\prime}n^{\prime} \atop mn}\left({\bf r}^{\prime}_1
\ {\bf r}^{\prime}_2 \atop 
      {\bf r}_1\ {\bf r}_2\right)$ is given in Appendix~A. 

Now we proceed to a more detailed derivation of the matrix representations of the exchange
interaction for cubic III-V semiconductors, such as GaAs, InAs and GaN,
 by taking into account the  conduction band~$\Gamma_6^c$, the 
heavy-hole and light-hole bands~$\Gamma_8^v$ and the  split-off
band~$\Gamma_7^v$. 
The Bloch functions for these bands and the time reversal of those for the
valence bands are given in Appendix~B.\cite{note1} 
 We further denote the conduction band Bloch functions as
\begin{equation}
  |c_1\rangle=|\frac{1}{2},\frac{1}{2}\rangle_c, \ \ \ \  |c_2\rangle=|\frac{1}{2},-\frac{1}{2}\rangle_c. 
\end{equation}
and the hole Bloch functions as:
\begin{eqnarray}
  |v_1\rangle=|\frac{3}{2},+\frac{3}{2}\rangle_h,\  
  |v_2\rangle=|\frac{3}{2},+\frac{1}{2}\rangle_h,\ 
  |v_3\rangle=|\frac{3}{2},-\frac{1}{2}\rangle_h,\\
  |v_4\rangle=|\frac{3}{2},-\frac{3}{2}\rangle_h,\ 
  |v_5\rangle=|\frac{1}{2},+\frac{1}{2}\rangle_h,\ 
  |v_6\rangle=|\frac{1}{2},-\frac{1}{2}\rangle_h.
\end{eqnarray}
  
We perform the Fourier expansion as $U({\bf r})= \frac{1}{8\pi^3}\int d{\bf q}
U_{q}e^{i{\bf q}\cdot{\bf r}}$ where $U_{q}=\frac{e^2}{\varepsilon_0\kappa
  q^2}$, and then the term of LR exchange Hamiltonian in 
Eq.~(\ref{equ:Ham_exchange_long}) transforms into the form
\begin{eqnarray}
 \nonumber
 &&  \hspace{-0.8cm}H^{{\rm LR}}_{m^{\prime}n^{\prime} \atop mn}\left({\bf r}^{\prime}_1\ {\bf r}^{\prime}_2 \atop {\bf r}_1\ {\bf
      r}_2\right)= \frac{1}{8\pi^3} \int d{\bf q}U_q\Big(\sum_{\alpha\beta}Q^{\alpha\beta}_{m^{\prime}\Theta n \atop
      \Theta n^{\prime}m}q_\alpha q_\beta \Big)\\
  && \ \ \ \ \ \mbox{}\times e^{i{\bf q}\cdot({\bf r}_1-{\bf
      r}^{\prime}_1)}\delta({\bf r}_1-{\bf r}_2)\delta({\bf r}^{\prime}_1-{\bf r}^{\prime}_2),
  \label{equ:Long2}
\end{eqnarray}
with $\Theta$ being the time reversal operator.
We define the band-relevant part of the LR exchange Hamiltonian as $Q_{m^{\prime}n^{\prime} \atop
  mn}({\bf q})$ with
\begin{equation}
  Q_{m^{\prime}n^{\prime} \atop mn}({\bf q})=\sum_{\alpha\beta}Q^{\alpha\beta}_{m^{\prime}\Theta n \atop
      \Theta n^{\prime}m}q_\alpha q_\beta.
\end{equation}
Then the parity of the Bloch functions enables us to write the matrix representation of
$Q_{m^{\prime}n^{\prime} \atop mn}({\bf q})$ in a somewhat simple way. With the basis
$|mn\rangle$ taken in the order $|c_1v_1\rangle, |c_2v_1\rangle, |c_1v_2\rangle, |c_2v_2\rangle,
  |c_1v_3\rangle, |c_2v_3\rangle,|c_1v_4\rangle,|c_2v_4\rangle, \\ |c_1v_5\rangle, |c_2v_5\rangle,
|c_1v_6\rangle$, $|c_2v_6\rangle$, $Q_{m^{\prime}n^{\prime} \atop mn}({\bf q})$
takes the form
\begin{widetext}
  \begin{eqnarray}
    \hspace{-0.45cm}Q_{m^{\prime}n^{\prime} \atop mn}({\bf q})=
    \setlength{\arraycolsep}{0.4mm}
    \left(\begin{array}{cccccccccccc}
        0&0&0&0&0&0&0&0&0&0&0&0\\
       &A{K^2}&{-\frac{AK^2}{\sqrt{3}}}&{-\frac{2AK_{-}q_z}{\sqrt{3}}}&{\frac{2AK_{-}q_z}{\sqrt{3}}}&{-\frac{AK_{-}^2}{\sqrt{3}}}&A{K_{-}^2}&0&{{\frac{-BK^2}{\sqrt{6}}}}&{{\frac{BK_{-}q_z}{\sqrt{6}}}}&{{\frac{BK_{-}q_z}{\sqrt{6}}}}&{{\frac{BK_{-}^2}{\sqrt{6}}}}\\
        &&{\frac{AK^2}{3}}&{\frac{2AK_{-}q_z}{3}}&{-\frac{2AK_{-}q_z}{3}}&{\frac{AK^2_{-}}{3}}&{-\frac{AK^2_{-}}{\sqrt{3}}}&0&{{\frac{BK^2}{3\sqrt{2}}}}&{{-\frac{BK_{-}q_z}{3\sqrt{2}}}}&{{-\frac{BK_{-}q_z}{3\sqrt{2}}}}&{{-\frac{BK_{-}^2}{3\sqrt{2}}}}\\
        &&&{\frac{4Aq_z^2}{3}}&{-\frac{4Aq_z^2}{3}}&{\frac{2AK_{-}q_z}{3}}&{-\frac{2AK_{-}q_z}{\sqrt{3}}}&0&{{\frac{\sqrt{2}BK_{+}q_z}{3}}}&{{-\frac{\sqrt{2}Bq_z^2}{3}}}&{{-\frac{\sqrt{2}Bq_z^2}{3}}}&{{-\frac{\sqrt{2}BK_{-}q_z}{3}}}\\
        &&&&{\frac{4Aq_z^2}{3}}&{-\frac{2AK_{-}q_z}{3}}&{\frac{2AK_{-}q_z}{\sqrt{3}}}&0&{{-\frac{\sqrt{2}BK_{+}q_z}{3}}}&{{\frac{\sqrt{2}Bq_z^2}{3}}}&{{\frac{\sqrt{2}Bq_z^2}{3}}}&{{\frac{\sqrt{2}BK_{-}q_z}{3}}}\\
        &&&&&{\frac{AK^2}{3}}&{-\frac{AK^2}{\sqrt{3}}}&0&{{\frac{BK_{+}^2}{3\sqrt{2}}}}&{-{\frac{BK_{+}q_z}{3\sqrt{2}}}}&{{-\frac{BK_{+}q_z}{3\sqrt{2}}}}&{{-\frac{BK^2}{3\sqrt{2}}}}\\
        &&&&&&A{K^2}&0&{{-\frac{BK_{+}^2}{\sqrt{6}}}}&{{\frac{BK_{+}q_z}{\sqrt{6}}}}&{{\frac{BK_{+}q_z}{\sqrt{6}}}}&{{\frac{BK^2}{\sqrt{6}}}}\\
        &&&&&&&0&0&0&0&0\\
        &&&&&&&&{{\frac{CK^2}{3}}}&{{-\frac{CK_{-}q_z}{3}}}&{{-\frac{CK_{-}q_z}{3}}}&{{-\frac{CK_{-}^2}{3}}}\\
        &&&&&&&&&{{\frac{Cq_z^2}{3}}}&{{\frac{Cq_z^2}{3}}}&{{\frac{CK_{\Red{-}}q_z}{3}}}\\
        &&&&&&&&&&{{\frac{Cq_z^2}{3}}}&{\frac{CK_{\Red{-}}q_z}{3}}\\
        &&&&&&&&&&&{{\frac{CK^2}{3}}}
      \end{array}\right).\label{equ:Matrix_long1}
 \end{eqnarray}
\end{widetext}
The other half of the matrix is obtained by taking the Hermitian conjugate. The
matrices below are given in the same way. Here $K_{\pm}=q_x\pm iq_y$ and ${\bf
  K}=(q_x,q_y)$. For the coefficients, $A=\frac{\hbar^2P^2}{2m_0^2E_g^2}$,  
$B=\frac{\hbar^2P^2}{m_0^2E_g(E_g+\Delta_{so})}$ and 
$C=\frac{\hbar^2P^2}{m_0^2(E_g+\Delta_{so})^2}$ with $m_0$, $E_g$ and
$\Delta_{so}$ standing for the free electron mass, the
band gap and the spin-orbit splitting, respectively.\cite{Winkler} 
$P=\left\langle S\left|p_x\right|X \right\rangle=\left\langle S\left|p_y\right|Y  
\right\rangle=\left\langle S\left|p_z\right|Z \right\rangle
$. $\frac{2P^2}{m_0}=E_P$ with $E_P$ being a band structure parameter in energy
unit.\cite{Vurga}  Approximation
has been made that the element $R=\frac{\hbar}{4mc^2}\left\langle S\left|\frac{\partial}{\partial
      x}V_0\right|X \right\rangle=\frac{\hbar}{4mc^2}\big\langle 
  S\big|\frac{\partial}{\partial y}V_0\big|Y
\big\rangle=\frac{\hbar}{4mc^2}\left\langle S\left|\frac{\partial}{\partial
      z}V_0\right|Z \right\rangle$ from the spin-orbit
  coupling in ${\bm \pi}$ [given in Eq.~(\ref{equ:Ham_val})], is
neglected since it is always much smaller than $P$.
The 12 $\times$ 12 matrix
representation of SR exchange interaction is written as 
  \begin{widetext}
 \begin{eqnarray}
 \nonumber
    &&\hspace{-2.3cm} H^{{\rm SR}}_{m^{\prime}n^{\prime} \atop mn}\left({\bf r}^{\prime}_1\ {\bf r}^{\prime}_2 \atop {\bf r}_1\ {\bf
      r}_2\right)=D\delta({\bf r}_1-{\bf r}_2)\delta({\bf r}_1-{\bf r}^{\prime}_1)\delta({\bf r}_2-{\bf r}^{\prime}_2)\\
    \setlength{\arraycolsep}{1mm}
 &&\hspace{-0.1cm}\times\left(\begin{array}{cccccccccccc}
     0&0&0&0&0&0&0&0&0&0&0&0\\
      &1&-\frac{1}{\sqrt{3}}&0&0&0&0&0&{-\sqrt{\frac{2}{3}}}&0&0&0\\
      &&\frac{1}{3}&0&0&0&0&0&{\frac{\sqrt{2}}{3}}&0&0&0\\
    &&&\frac{2}{3}&-\frac{2}{3}&0&0&0&0&{-\frac{\sqrt{2}}{3}}&{-\frac{\sqrt{2}}{3}}&0\\
    &&&&\frac{2}{3}&0&0&0&0&{\frac{\sqrt{2}}{3}}&{\frac{\sqrt{2}}{3}}&0\\
    &&&&&\frac{1}{3}&-\frac{1}{\sqrt{3}}&0&0&0&0&{-\frac{\sqrt{2}}{3}}\\
    &&&&&&1&0&0&0&0&{\sqrt{\frac{2}{3}}}\\
    &&&&&&&0&0&0&0&0\\
    &&&&&&&&{\frac{2}{3}}&0&0&0\\
    &&&&&&&&&{\frac{1}{3}}&{\frac{1}{3}}&0\\
    &&&&&&&&&&{\frac{1}{3}}&0\\
    &&&&&&&&&&&{\frac{2}{3}}
 \end{array}\right), \label{equ:Matrix_short1}
\end{eqnarray}
  \end{widetext}
in which
\begin{equation}
    D= \frac{1}{V}\frac{1}{8 \pi^3} \int
  d{\bf q} U_q |\langle S|e^{i{\bf q}\cdot{\bf r}}|X \rangle|^2.
\end{equation}
An $8\times 8$ matrix representation of the LR
  and SR exchange interactions can be found 
in Refs.~\onlinecite{Maialle} and \onlinecite{Taka2}, where only the heavy- and
light-hole bands are included. They are the same as the $8\times
  8$ submatrices at the top left corner of our expressions, but in two-\cite{Maialle} or
  zero-dimensional\cite{Taka2} forms. Whereas in  
Ref.~\onlinecite{Kada}, a $4\times 4$ exciton Hamiltonian was presented
including only the heavy-hole band. So the previous expressions correspond to only part of
our results, where the effects from the split-off band and even the
light-hole band are absent. 
In comparison with the matrix form of exchange interaction Hamiltonian in Ref.~\onlinecite{Maialle}, we obtain
\begin{eqnarray}\Delta E_{\rm LT}&=&
  \frac{2e^2\hbar^2E_P}{3\pi\epsilon_0\kappa m_0a_{\rm Bohr}^3E_g^2},\\
  D&=&\frac{3}{4}\pi a_{\rm Bohr}^3\Delta E_{\rm SR}, \label{equ:DD} 
 \label{equ:bohr}  
\end{eqnarray} 
where $E_g$ is the band gap, $\Delta E_{\rm LT}$ and $\Delta E_{\rm SR}$ are the longitudinal-transverse
and the singlet-triplet
splittings in bulk which can be obtained by
experiment.\cite{Andreani,Ekardt} $m_e^\ast$ is the effective mass of the
conduction electron, $\gamma_1$ is the band parameter and 
$a_{\rm Bohr}=4\pi\epsilon_0\kappa\hbar^2\mu/e^2$ with 
$\mu^{-1}=1/m_e^\ast+\gamma_1/m_0$ is
the exciton Bohr radius in bulk.\cite{Ekardt}

\subsubsection{e-h exchange interaction in  QW }
With the e-h exchange Hamiltonian described above, one can investigate the
properties of excitons by explicitly including the contributions of 
the heavy-hole, light-hole and split-off bands. In QWs with small well
width, as an usual procedure, two-dimensional (2D) exciton bound
states are used to approximately count
the effect of direct Coulomb interaction and 
the e-h exchange interaction is treated perturbatively.\cite{Chen, Maialle} 
For the QD system in the strong-confinement regime, in the literature, people first solve
the confinement and then treat the direct Coulomb and e-h exchange
interactions as perturbation.\cite{Ivchenko3,Horo} However, in QWs or QDs 
the characteristic size of which is comparable or even larger than the exciton Bohr
radius~$a_{\rm Bohr}$, which means that the Coulomb interaction tends to overtake the
confinement, one has to solve the Schr\"odinger equation with both the
confinement potential and the direct Coulomb interaction included. In this
case, it can be extremely CPU expensive to employ the 12$\times$12 e-h exchange 
Hamiltonian. So the L\"owdin partitioning method\cite{Winkler, Lowdin, Luttinger2} is
employed to derive the e-h exchange interaction in a smaller Hilbert space, while
still taking into account the confinement-induced valence band mixing.

We start from a system with strong confinement along the
$z$ direction (i.e., the [001] direction). 
The infinite square well potential is employed. The QW width is
denoted as $l_z$. For the envelope function in the $z$ direction, only the lowest
subband is relevant for both electron and hole.

It is noted that the hole Hamiltonian is not described by the so-called Luttinger
Hamiltonian itself,\cite{Luttinger1} which is the Hamiltonian for valence
electrons.\cite{Cho2} One has to transform it to hole space according to rules given in
Refs.~\onlinecite{Bir1} and \onlinecite{Bir2}. 
The obtained hole Hamiltonian in the valence bands $\Gamma_7^v$ and
$\Gamma_8^v$ takes the form as $-1$ times the 6$\times$6 Luttinger 
Hamiltonian.\cite{Winkler,Luttinger1}
After applying a strong confinement along the $z$ direction,
the Luttinger Hamiltonian is deduced to 2D form where the odd terms of $k_z$ vanish and
$\langle k_z^2\rangle=\pi^2/l_z^2$. The values of the minima of the diagonal
elements for the heavy, light and split-off holes are now separated due to their different effective masses
in the $z$ direction.\cite{Winkler} The heavy-hole band lies 
energetically much lower than the other two. 
This enables us to apply the L\"owdin
partitioning\cite{Winkler,Lowdin, Luttinger2} and get decoupled new basis
functions for holes. The lowest subbands are heavy-hole-like states, which are 
admixtures of the heavy- with the light- and split-off-hole states.

 The L\"owdin transformation of the hole Hamiltonian is given by $\tilde H^h = 
e^{-S}H^he^S$ where $S$ is an anti-Hermitian 6$\times$6 matrix. The
basis functions transform as $\tilde{\psi}_n=\sum_m(e^S)_{mn}\psi_m$. Up to the 
first-order approximation, one obtains
$\tilde{\psi}_n=\sum_m(\delta_{mn}+S_{mn})\psi_m$, with   
$\delta_{mn}$ being the Kronecker delta.  
Up to the second order, the effective Hamiltonian of the new 
heavy-hole-like subbands takes the form 
\begin{eqnarray}
 && \tilde H^{\rm HH} = 
    \setlength{\arraycolsep}{1mm}
 \left(\begin{array}{cc}
   h&0\\
   0&h
 \end{array}\right), \\
 &&h=\frac{\hbar^2}{2m_0}(\gamma_1+\gamma_2)k_{\parallel}^2+\frac{\hbar^2\pi^2}{2m_0l_z^2}(\gamma_1-2\gamma_2),
 \label{equ:Ham_hole}
 \end{eqnarray}
with $k_{\parallel}=(k_x,k_y)$, $k_x=-i\partial_x$ and
$k_y=-i\partial_y$.
The non-zero elements of the matrix $S$ up to the first order read
\begin{eqnarray}
 S_{31}&=&\frac{1}{E_{ab}}\frac{\hbar^2}{2m_0}(\sqrt{3}\gamma_2K +
 i2\sqrt{3}\gamma_3k_xk_y),\label{equ:S31}\\
 S_{61}&=&-\frac{1}{E_{ac}}\frac{\hbar^2}{2m_0}(\sqrt{6}\gamma_2K +
 i2\sqrt{6}\gamma_3k_xk_y),\\
S_{24}&=&\frac{1}{E_{ab}}\frac{\hbar^2}{2m_0}(\sqrt{3}\gamma_2K -
 i2\sqrt{3}\gamma_3k_xk_y),\\
S_{54}&=&\frac{1}{E_{ac}}\frac{\hbar^2}{2m_0}(\sqrt{6}\gamma_2K -
 i2\sqrt{6}\gamma_3k_xk_y).\label{equ:S54}
 \end{eqnarray}
where
\begin{eqnarray}
  &&K=k_x^2-k_y^2, \ \ 
  E_{ab}=\frac{\hbar^2}{2m_0}4\gamma_2k_{z}^2, \\
  &&E_{ac}=\frac{\hbar^2}{2m_0}(2\gamma_2-\gamma_1+\frac{m_0}{m_{so}})k_{z}^2+\Delta_{so}.
  \end{eqnarray}
Here $\gamma_i$ are the band parameters\cite{Winkler} and $E_{ab}$~($E_{ac}$) stands for the energy
splitting between the heavy-hole and light-hole (split-off) subbands.
The other half of $S$ is obtained from the relation $S^\dagger=-S$.
New Bloch functions for the heavy-hole-like states become
\begin{eqnarray}
  |V_1\rangle &=& |v_1\rangle + S_{31}|v_3\rangle + S_{61}|v_6\rangle,\label{equ:V1}\\
  |V_2\rangle &=& |v_4\rangle + S_{24}|v_2\rangle + S_{54}|v_5\rangle.\label{equ:V2}
  \end{eqnarray}
Since the major component of $|V_1\rangle$ ($|V_2\rangle$) is 
$|v_1\rangle$ ($|v_4\rangle$),  we still denote the coupled
states $|V_1\rangle$ and $|V_2\rangle$ as spin $\pm \frac{3}{2}$ states for
simplicity in the following. The new ``heavy-hole'' bands are now admixtures of
the heavy-hole, light-hole and split-off bands. 
It is this band-mixing effect that makes the dark
exciton states, constructed as $|c_1V_1\rangle$ or $|c_2V_2\rangle$, become partially
optically allowed.\cite{Bayer1,Bayer3,Gorycal,Nirmal}

Now we are ready to derive the e-h exchange interaction in 4$\times$4 matrix
representation with $|c_iV_j\rangle$ as the new Bloch wave function from 
Eqs.~(\ref{equ:Matrix_long1}) and (\ref{equ:Matrix_short1}). In order 
to write the Hamiltonian in a simple way, we note that since
\begin{equation}
  \nonumber
  \langle x^{\prime}y^{\prime},n_{z1}|\hat{A} | xy,n_{z2}\rangle=\langle
  n_{z1}|z^{\prime}\rangle\langle x^{\prime}y^{\prime},z^{\prime}|\hat{A} |xy,z\rangle\langle z | n_{z2}\rangle,
  \end{equation}
where $\hat{A}$ is an arbitrary operator and the Einstein summation convention
is presumed, we can always write the
exchange Hamiltonian in three-dimensional (3D) space while keeping in mind that the formulae hold true only
for QW or QD systems with strong confinement in one direction. 

The LR exchange interaction, given in the basis taken in the order
$|c_1V_1\rangle, |c_2V_1\rangle, |c_1V_2\rangle, |c_2V_2\rangle$ (or 
      expressed as the eigenstates of $J_z$: $|+2\rangle$, $|+1\rangle$,
 $|-1\rangle$, $|-2\rangle$), is written as
\begin{eqnarray}
    \hspace{-0.45cm}\tilde{H}^{{\rm LR}}_{m^{\prime}n^{\prime} \atop mn}\left({\bf r}^{\prime}_1\ {\bf r}^{\prime}_2 \atop {\bf r}_1\ {\bf
      r}_2\right)=
    \setlength{\arraycolsep}{1.5mm}
 \left(\begin{array}{cccc}
   0&0&0&0\\
   &\tilde{H}_{22}^{\rm LR}&\tilde{H}_{23}^{\rm LR}&0\\
   &&\tilde{H}_{33}^{\rm LR}&0\\
   &&&0
 \end{array}\right),
\label{equ:Ham_long}
\end{eqnarray}
  in which
  \begin{widetext}
  \begin{eqnarray}
  \nonumber
    \tilde{H}_{22}^{\rm LR}&=&\frac{\hbar^2P^2}{8\pi^3m_0^2E_g^2}\int d{\bf q}U_{{\bf q}}
    \left\{e^{i{\bf q}\cdot({\bf r}_1-{\bf r}^{\prime}_2)}\Big[
      \frac{1}{2}(q_x^2+q_y^2)-\frac{1}{\sqrt{2}}(q_x\Red{-}iq_y)^2\Big(\frac{S^{\ast}_{31}}{\sqrt{6}}-\frac{S^{\ast}_{61}}{\sqrt{3}}\Big)\Big]\right. \\ &&\left.\mbox{}
    -e^{i{\bf q}\cdot({\bf r}^{\prime}_1-{\bf r}_2)}
      \Big[\frac{1}{\sqrt{2}}(q_x\Red{+}iq_y)^2\Big(\frac{S^{\prime}_{31}}{\sqrt{6}}-\frac{S^{\prime}_{61}}{\sqrt{3}}\Big)\Big]\right\} 
      \delta({\bf r}_1-{\bf r}_2)\delta({\bf r}^{\prime}_1-{\bf r}^{\prime}_2),\label{equ:Ham_long22}\\
  \nonumber
     \tilde{H}_{23}^{\rm LR}&=&\frac{\hbar^2P^2}{8\pi^3m_0^2E_g^2}\int d{\bf q}U_{{\bf q}}
      \left\{e^{i{\bf q}\cdot({\bf r}_1-{\bf r}^{\prime}_2)}\Big[
      \frac{1}{2}(q_x-iq_y)^2-\frac{1}{\sqrt{2}}(q_x^2+q_y^2)\Big(\frac{S_{31}}{\sqrt{6}}-\frac{S_{61}}{\sqrt{3}}\Big)\Big]\right.\\
    &&\left.\mbox{}-e^{i{\bf q}\cdot({\bf r}^{\prime}_1-{\bf
          r}_2)}\Big[\frac{1}{\sqrt{2}}(q_x^2+q_y^2)\Big(\frac{S^{\prime \Red{\ast}
          }_{24}}{\sqrt{6}}+\frac{S^{\prime \Red{\ast}}_{54}}{\sqrt{3}}\Big)\Big]\right\}\delta({\bf
        r}_1-{\bf r}_2)\delta({\bf r}^{\prime}_1-{\bf r}^{\prime}_2),\label{equ:Ham_long23}\\
 \nonumber
      \tilde{H}_{33}^{\rm LR}&=&\frac{\hbar^2P^2}{8\pi^3m_0^2E_g^2}\int d{\bf q}U_{{\bf q}}
    \left\{e^{i{\bf q}\cdot({\bf r}_1-{\bf r}^{\prime}_2)}\Big[
      \frac{1}{2}(q_x^2+q_y^2)-\frac{1}{\sqrt{2}}(q_x\Red{+}iq_y)^2\Big(\frac{S^{\ast}_{24}}{\sqrt{6}}+\frac{S^{\ast}_{54}}{\sqrt{3}}\Big)
      \right.\Big]\\ && \left.\mbox{}
    -e^{i{\bf q}\cdot({\bf r}^{\prime}_1-{\bf r}_2)}
      \Big[ \frac{1}{\sqrt{2}}(q_x\Red{-}iq_y)^2\Big(\frac{S^{\prime
         }_{24}}{\sqrt{6}}+\frac{S^{\prime}_{54}}{\sqrt{3}}\Big)\Big]\right\}
      \delta({\bf r}_1-{\bf r}_2)\delta({\bf r}^{\prime}_1-{\bf r}^{\prime}_2).\label{equ:Ham_long33}
  \end{eqnarray}
  \end{widetext}
Here $S^{\ast}_{ij}$ is the complex conjugate of $S_{ij}$ defined in
 Eqs.~(\ref{equ:S31})-(\ref{equ:S54}) with ${\bf k}\rightarrow {\bf k}_2$, and
$S^{\prime}_{ij}$ is the same as $S_{ij}$ but with ${\bf k}\rightarrow {\bf
  k}^{\prime}_2$. 
The matrix of short-range exchange interaction is given in the same way:
  \begin{eqnarray}
    \hspace{-0.45cm}\tilde{H}^{{\rm SR}}_{m^{\prime}n^{\prime} \atop mn}\left({\bf r}^{\prime}_1\ {\bf r}^{\prime}_2 \atop {\bf r}_1\ {\bf
      r}_2\right)=
    \setlength{\arraycolsep}{1.5mm}
 \left(\begin{array}{cccc}
   0&0&0&0\\
   &\tilde{H}_{22}^{\rm SR}&\tilde{H}_{23}^{\rm SR}&0\\
   &&\tilde{H}_{33}^{\rm SR}&0\\
   &&&0
 \end{array}\right),\label{equ:Ham_ss}
\end{eqnarray}
in which
  \begin{eqnarray}
    \hspace{-0.5cm}\tilde{H}_{22}^{\rm SR}&=&D\delta({\bf r}_1-{\bf r}_2)\delta({\bf r}_1-{\bf
      r}^{\prime}_1)\delta({\bf r}_2-{\bf r}^{\prime}_2),\label{equ:Ham_short22}\\
  \nonumber
    \hspace{-0.5cm}\tilde{H}_{23}^{\rm
      SR}&=&-\sqrt{2}D\Big(\frac{S^{\Red{\prime}}_{31}}{\sqrt{6}}-\frac{S^{\Red{\prime}}_{61}}{\sqrt{3}}\Red{+}\frac{S^{\ast}_{24}}{\sqrt{6}}\Red{+}\frac{S^{\ast}_{54}}{\sqrt{3}}\Big)\\ 
    &&
    \mbox{}\times\delta({\bf r}_1-{\bf r}_2)\delta({\bf r}_1-{\bf
      r}^{\prime}_1)\delta({\bf r}_2-{\bf r}^{\prime}_2),\label{equ:Ham_short23}\\
    \hspace{-0.5cm}\tilde{H}_{33}^{\rm SR}&=&D\delta({\bf r}_1-{\bf r}_2)\delta({\bf r}_1-{\bf
      r}^{\prime}_1)\delta({\bf r}_2-{\bf r}^{\prime}_2).\label{equ:Ham_short33}
 \end{eqnarray}

From the above expressions, one
 can see that the four-fold-degenerate exciton 
states are split when the e-h exchange
  interaction is taken into account. When the 
diagonal matrix elements are included, the
 quadruplet is split into two doublets: the bright and dark doublets with the
 bright one lying above the dark one. 
The off-diagonal matrix elements further couple
the two bright exciton states together and cause the doublet splitting. 
Moreover, it is noted that in Eqs.~(\ref{equ:Ham_long22})-(\ref{equ:Ham_long33})
and (\ref{equ:Ham_short22})-(\ref{equ:Ham_short33}) the terms without $S^\ast_{ij}$
and $S^{\prime}_{ij}$ are derived from the heavy-hole band; the terms with $S^\ast_{31(24)}$
and $S^{\prime}_{31(24)}$ are derived from the light-hole band and the terms with $S^\ast_{61(54)}$
and $S^{\prime}_{61(54)}$ are derived from the split-off band.  
It is noted that, as shown in Eq.~(\ref{equ:Ham_ss}), 
the SR exchange interaction now
directly couples the $J_z=\pm 1$ exciton states due to the 
confinement-induced band mixing. This coupling is missing when 
only the heavy-hole band is included as in Ref.~\onlinecite{Kada}.

We note that the exciton wave function $\Psi({\bf r}_1,{\bf r}_2)$ in
  Eq.~(\ref{equ:Wavef1}) is actually 
the same as Eq.~(2.1) in Ref.~\onlinecite{Taka2} up to a conventional constant, while the latter
is written in the framework of second-quantization. This brings some insights into
how the electron and hole share the properties of identical particles. 
The exciton wave function Eq.~(\ref{equ:Wavef1}) is also widely used as a
truncated one\cite{Efros,Ekardt,Chen,Romestain} 
\begin{equation}
  \Psi({\bf r}_e,{\bf r}_h) = \sum_{cv}F_{cv}({\bf r}_e,{\bf r}_h)\psi_{c{\bf
      k}_0}({\bf r}_e)\tilde{\psi}_{v{\bf k}_0}({\bf r}_h),
\label{equ:Wavef2}
\end{equation}
 which is sufficient to
describe most of the properties of exciton, e.g., spin dynamics\cite{Maialle} and
fine structure.\cite{Efros,Chen}
 We use this expression in the
calculation of the exciton spin relaxation rate.

\subsection{Scaling of the e-h exchange interaction}
The study of size scaling of the e-h exchange interaction in semiconductor QDs
helps to understand the experimental results\cite{Zunger3,Norris,Nirmal} and
provides an intuitive understanding of how it varies with 
dot size.\cite{Kada,Taka2,Zunger4,Zunger5,Efros,Chamarro,Leung} 
The scaling rules were established in the strong confinement regime.\cite{Kada,Taka2,Romestain,Efros} 
For the doublet splitting, Takagahara\cite{Taka2} and Kadantsev and
Hawrylak\cite{Kada} established that the LR exchange interaction which
determines the splitting scales as $1/L^3$ with $L$ standing for the
characteristic size of the QD. Their numerical results 
(fit to $C/L^n$ with $n$$\approx$$1.3$ in
Ref.~\onlinecite{Taka2} and $n=1.3$$\sim$$1.5$ in Ref.~\onlinecite{Kada}) showed
discrepancy from the $1/L^3$ dependence and the discrepancy was attributed to
the details of the envelope functions.  
For the BD exchange splitting, most works were carried out retaining only the SR exchange
interaction and therefore the BD splitting was assumed to scale as
$1/L^3$.\cite{Romestain,Efros,Nirmal,Leung} However, clear deviation of
the experimental results from the
$1/L^3$ law was reported in Ref.~\onlinecite{Chamarro}. Moreover, by fitting the
size dependence of the numerical results to the scaling law $1/L^n$, Franceschetti {\em
  et al}.\cite{Zunger4} reached $n=1.93$ for InP
nanocrystals and $n=1.97$ for CdSe nanocrystals, whereas in
Ref.~\onlinecite{Zunger5}, $n=2.51$ was obtained for Si QDs. This discrepancy to the $1/L^3$
law was attributed to the presence of the LR component of the e-h exchange
interaction. Obvious confusion is seen and hence a reexamination is necessary.

We point out that not only the values of the exchange splittings, but also the
scaling laws, depend on the dot size. So the investigation of
size scaling of the e-h exchange interaction is carried out in two limits: the
strong and weak confinement limits. 
Moreover, 3D and 2D scalings are performed. For the 3D scaling, 
the dot size is varied in all three dimensions; whereas for the 2D scaling, the dot
height is fixed and only the lateral size is varied.  The characteristic
size of the variation is denoted as $L$ and the dot height $l_z$ in the 2D
scaling is fixed and assumed to be much smaller than the
exciton Bohr radius.

Since the leading terms of the e-h exchange interaction originate from the
heavy-hole-exciton basis in cubic 
III-V semiconductor QDs (e.g., GaAs, InAs and GaN QDs investigated in this
paper, shown in Sec.~III), we focus on these terms in
Eqs.~(\ref{equ:Ham_long})-(\ref{equ:Ham_short33}), i.e., terms without
$S^\ast_{ij}$ and $S^{\prime}_{ij}$. Then the first terms in 
Eqs.~(\ref{equ:Ham_long22})-(\ref{equ:Ham_long33}) scale as 
\begin{equation}
  \int d{\bf q}U_{{\bf q}}e^{i{\bf q}\cdot({\bf r}_1-{\bf r}^{\prime}_2)}q_iq_j
  \propto \partial_i\partial_j\frac{e^2}{4\pi\epsilon_0\kappa|{\bf r}^{\prime}_1-{\bf
      r}_2|} \propto \frac{1}{L^3}. \label{equ:scaling-1}
\end{equation}
For the strong confinement limit where $L$ is much 
smaller than the exciton Bohr radius, the exciton envelop function $F_{cv}({\bf
  r}_1,{\bf r}_2)$ scales as $\propto \frac{1}{L^3}$ in the 3D scaling and $\propto
\frac{1}{L^2}$ in the 2D scaling due to the normalization 
condition.\cite{Taka2} In the calculation of matrix elements in
Eq.~(\ref{equ:SE1}), we reach the scaling laws for the LR exchange terms
\begin{eqnarray}
 && L^3L^3\frac{1}{L^3}\frac{1}{L^3}\frac{1}{L^3}\propto\frac{1}{L^3} \ \  ({\rm 3D}),\label{equ:scaling-L1}\\
 && L^2L^2\frac{1}{L^3}\frac{1}{L^2}\frac{1}{L^2}\propto\frac{1}{L^3}\ \  ({\rm
   2D}), \label{equ:scaling-L2} 
\end{eqnarray}
and for the short-range exchange terms 
\begin{eqnarray}
  && L^3\frac{1}{L^3}\frac{1}{L^3}\propto\frac{1}{L^3}\ \   ({\rm 3D}),\label{equ:scaling-S1}\\   
  && L^2\frac{1}{L^2}\frac{1}{L^2}\propto\frac{1}{L^2}\ \  ({\rm 2D}).  \label{equ:scaling-S2}
\end{eqnarray}
These are the same as those in Ref.~\onlinecite{Taka2}.

However, for the
weak confinement limit where $L$ is much larger than the exciton Bohr radius, the
relative motion of electron and hole is not sensitive to the value of $L$. Only the motion
of the center-of-mass of the exciton is affected by the confinement. Therefore the scaling
of the exciton envelope function changes to $F_{cv}({\bf r}_1,{\bf r}_2)\propto
\frac{1}{L^2}$ in the 3D scaling and $\propto \frac{1}{L}$
in the 2D scaling. Here we have assumed that
$l_z$ is still smaller than the exciton Bohr radius in the weak confinement
limit under  consideration. Now terms of  the LR exchange interaction scale as
\begin{eqnarray}
 && L^3L^3\frac{1}{L^3}\frac{1}{L^2}\frac{1}{L^2}\propto\frac{1}{L} \ \  ({\rm 3D}),\label{equ:scaling-L3}\\
 && L^2L^2\frac{1}{L^3}\frac{1}{L}\frac{1}{L}\propto\frac{1}{L}\ \  ({\rm
   2D}), \label{equ:scaling-L4} 
\end{eqnarray}
The short-range exchange terms scale as 
\begin{equation}
 L^3\frac{1}{L^2}\frac{1}{L^2}\propto\frac{1}{L}\   ({\rm 3D}),\ \   
 L^2\frac{1}{L}\frac{1}{L}\propto\frac{1}{L^0}\   ({\rm 2D}), \label{equ:scaling-S5}
\end{equation}

For genuine situation between these two limits, terms of  the LR exchange interaction scale in the range
of $\frac{1}{L} \sim \frac{1}{L^3}$ in both 3D and 2D scalings 
while those of the short-range exchange 
interaction scale in the range of $\frac{1}{L}\sim\frac{1}{L^3}$ 
in the 3D scaling and
$\frac{1}{L^0}\sim\frac{1}{L^2}$ in the 2D scaling, respectively. In this way, we
are able to explain the confusion discussed at the beginning of this subsection:
the scaling of the e-h exchange interaction was investigated in different 
regimes of the confinement strength.
We will further check
these scaling rules in the next section in the analysis of the
 numerical results.

\subsection{Exciton spin relaxation}
The spin relaxation between the lowest two linear polarized exciton states $|X\rangle$
and $|Y\rangle$ induced by the asymmetry of QD determines the dynamics of the optical linear
polarization decay. The $|X\rangle$ and $|Y\rangle$ states are defined as
$|X\rangle=(|$+$1\rangle+|$$-$1$\rangle)/\sqrt{2}$ and
$|Y\rangle=-i(|$+$1\rangle-|$$-$$1\rangle)/\sqrt{2}$ where $|\pm$$1\rangle$ denotes the
optically active exciton states with total angular momentum in the $z$-direction
$J_z=\pm 1$.\cite{Marie} In GaN QDs, the effect
of surface roughness can be suppressed experimentally to such an extent that it can
be ignored. Therefore, the exciton spin relaxation is mainly assisted
by electron- and hole-phonon interactions induced by deformation potential and
piezoelectric field. From the Fermi
golden rule, the phonon-assisted relaxation rate from $|i\rangle$ to
$|f\rangle$  can be calculated by
\begin{eqnarray}
    \nonumber
  \Gamma_{i\to f}&=&\frac{2\pi}{\hbar}\sum_{{\bf q}\lambda}|M_{{\bf
      q}\lambda}|^2|\langle f|\chi|i\rangle|^2[\bar{n}_{{\bf
      q}\lambda}\delta(\epsilon_f-\epsilon_i-\hbar_{{\bf q}\lambda}) \\
  && \mbox{}+(\bar{n}_{{\bf
      q}\lambda}+1)\delta(\epsilon_f-\epsilon_i+\hbar_{{\bf q}\lambda})],\label{equ:relax}
  \end{eqnarray}
in which $M_{{\bf q}\lambda}$ and $\chi=e^{i{\bf q}\cdot{\bf r}_1}+e^{i{\bf
    q}\cdot{\bf r}_2}$ come from the electron- and hole-phonon interaction. $\bar{n}_{{\bf 
    q}\lambda}$ is the Bose distribution of phonon with mode $\lambda$ and
wave vector ${\bf q}$. In our calculation, the temperature is fixed at 0~K, so the
phonon absorption process is absent. 

In this paper, we take into account
 the electron- and hole-acoustic-phonon scattering due to the deformation
potential with $|M_{{\bf
    q}sl}|^2=\hbar\Xi^2q/(2\rho v_{sl})$, and due to the piezoelectric coupling with $|M_{{\bf
    q}sl}|^2=288\hbar\pi^2e^2e_{14}^2(q_xq_yq_z)^2/(\kappa^2\rho v_{sl}q^7)$
for the longitudinal mode and $\sum_{j=1,2}|M_{{\bf
    q}pt_j}|^2=32\hbar\pi^2e^2e_{14}^2[q_x^2q_y^2+q_y^2q_z^2
+q_z^2q_x^2-(3q_xq_yq_z)^2/q^2]/(\kappa^2\rho v_{st}q^5)$ for the two transverse
modes.\cite{Vogl,Lei,Wu} Here $\Xi$ 
, $\rho$, $e_{14}$ and $\kappa$ stand for the acoustic deformation potential, the volume density of
the material, the piezoelectric coupling constant and
 the static dielectric constant, respectively. $v_{sl}$~($v_{st}$) is the
longitudinal~(transverse) sound velocity.
Their values are given in Table~I.

\subsection{Numerical scheme}
In the computation we employ a disk-like QD model to simulate the real
QDs, with infinite square-well potential in the $z$-direction (i.e., the [001]
direction) and anisotropic
coaxial harmonic-oscillator potential as 
the in-plane confinement:\cite{Raymond,Hawrylak}
\begin{eqnarray}
  V_{e(h)}(z)&=&\left\{
    \begin{array}{ll}
     0,& 0<z<l_z \label{equ:con1}\\
      \infty,& {\rm others }
    \end{array}
  \right.,\\
    V_{e(h)}(x,y)&=& \frac{1}{2}m^\ast_{e(h){\parallel}}(\omega_{xe(h)}^2x^2 + \omega_{ye(h)}^2y^2),\label{equ:con2}
  \end{eqnarray}
where $\omega_{xe(h)}=\frac{\hbar}{m^\ast_{e(h)\parallel}l_x^2}$ and
$\omega_{ye(h)}=\frac{\hbar}{m^\ast_{e(h){\parallel}}l_y^2}$ with
$m^\ast_{e(h){\parallel}}$ denoting the effective mass of the 
electron (hole) in the plane. $l_x$ and $l_y$ are the characteristic lengths
of the harmonic-oscillator potentials along the $x$- and
$y$-directions and correspond to the major and/or minor diameters of the elliptic QD
in the plane. $l_z$ corresponds to the dot height. In our model the
single electron and hole experience different 
in-plane potentials but share the same confinement length. 
We adjust the relative magnitudes of $l_x$ and $l_y$ to control the anisotropy
of the QD and the magnitudes of $l_x$, $l_y$ and $l_z$ to vary the strength of
the confinement. 

\begin{widetext}
\begin{center}
\begin{table}[htb]
\caption{Material parameters used in the calculation (from
  Ref.~\onlinecite{Vurga} unless otherwise specified).}
\begin{tabular}{ccccccccc}
\hline\hline
&$E_g$ (eV)&$E_p$ (eV)&$m_e^\ast/m_0$&$m_{so}^\ast/m_0$&$\gamma_1$&$\gamma_2$&$\kappa$&$\varXi$ (eV)\\
GaN  &3.299&25.0&0.15&0.29&2.67&0.925$^{\rm e}$&9.7$^{\rm a}$&8.3$^{\rm d}$\\
GaAs &1.519&28.8&0.0665&0.172&6.85&2.5$^{\rm e}$&12.53$^{\rm a}$&8.5$^{\rm b}$\\
InAs &0.414&21.5&0.023&0.14&20.4&8.7$^{\rm e}$&15.15$^{\rm a}$&5.8$^{\rm b}$\\
\hline
&\mbox{}\mbox{}\mbox{}$\Delta_{\rm SO}$ (eV)\mbox{}\mbox{}\mbox{}&\mbox{}\mbox{}\mbox{}$\rho$
($10^3$~kg/m$^{3}$)\mbox{}\mbox{}\mbox{}&\mbox{}\mbox{}\mbox{}$e_{14}$~($10^8$~V/m)\mbox{}\mbox{}\mbox{}&\mbox{}\mbox{}\mbox{}$\Delta
E_{\rm LT}$ ($\mu$eV)\mbox{}\mbox{}\mbox{}&\mbox{}\mbox{}\mbox{}$\Delta E_{\rm SR}$
($\mu$eV)\mbox{}\mbox{}\mbox{}&\mbox{}\mbox{}\mbox{}$v_{sl}$ ($10^3$~m/s)\mbox{}\mbox{}\mbox{}&\mbox{}\mbox{}\mbox{}$v_{st}$ ($10^3$~m/s)\mbox{}\mbox{}\mbox{}&\mbox{}\mbox{}\mbox{}\mbox{}\mbox{} \\
GaN  &0.017&6.095$^{\rm d}$&43$^{\rm d}$&---&---&6.56$^{\rm d}$&2.68$^{\rm d}$&\\
GaAs &0.341&5.31$^{\rm b}$&14.1$^{\rm b}$&80$^{\rm b}$&20$^{\rm b}$&5.29$^{\rm b}$&2.48$^{\rm b}$&\\
InAs &0.38&5.9$^{\rm b}$&3.5$^{\rm b}$&---&0.3$^{\rm c}$&4.28$^{\rm b}$&1.83$^{\rm b}$&\\
\hline\hline
\end{tabular} \\
$^{\rm a}$ Ref.~\onlinecite{Jiang},\quad\quad
$^{\rm b}$ Ref.~\onlinecite{Landolt},\quad\quad
$^{\rm c}$ Ref.~\onlinecite{Kalt1},\quad\quad
$^{\rm d}$ Ref.~\onlinecite{Weng},\quad\quad\quad\quad\quad\quad\quad\quad\quad\quad\quad\quad\quad\quad\quad\quad\quad\quad\quad\quad\quad\quad\quad\quad\quad\quad\\
\quad\quad\quad\quad
$^{\rm e}$ Obtained from Ref.~\onlinecite{Vurga} by $\frac{1}{2}(\gamma_2+\gamma_3)$
in the spherical approximation.\cite{Winkler} $\gamma_3=\gamma_2$ in this paper.\hfill
\label{table1}
\end{table}
\end{center}
\end{widetext}

The eigen equation for the envelope function $F_{mn}({\bf r}_1,{\bf r}_2)$ [Eq.~(\ref{equ:SE1})] is 
solved by the exact diagonalization method. 
The total Hamiltonian [Eq.~(\ref{equ:Ham_total})] is separated into two parts: $H^{eh}=H_0+H^{\prime}$ where $H_0$
[Eq.~(\ref{equ:H0})] is diagonal in the real
space and easy to be solved,  and $H^{\prime}$ is the remaining parts of $H^{eh}$
which 
include the e-h exchange interaction
[Eqs.~(\ref{equ:Ham_long})-(\ref{equ:Ham_short33})] and $\tilde{H}^{\prime}$ given in 
Eq.~(\ref{equ:H_per}). The eigenfunctions of $H_0$ are taken
as the basis functions, and the total Hamiltonian is diagonalized in the Hilbert
space spanned by them. Detailed procedures
are laid out in Appendix~C. 

We stress that the direct Coulomb interaction is included in $H_0$, together with
the confinement,  and solved exactly. 
As pointed out in the Introduction, the direct Coulomb interaction is
comparable or even stronger than the lateral confinement of the QDs 
  under investigation.\cite{note2} So it
is not appropriate to construct the basis of exciton envelope functions by the product of single-particle wave functions of electron and hole\cite{Kada,Ivchenko3} and treat the
direct Coulomb interaction perturbatively. As we will show below, the strength 
of the direct Coulomb interaction strongly affects the calculated fine structure
splittings. Moreover, the big exciton binding energy obtained\cite{note3} indicates that the
exciton wave functions, as well as the the spectrum, are markedly modulated by the
direct Coulomb interaction. Therefore, the unperturbative treatment of the direct Coulomb
interaction is crucial to fully take into account its effect on the exciton fine
structures.

\section{Exciton fine structure in QDs}
\begin{figure}[htb]
  \begin{center}
    \includegraphics[width=8.5cm]{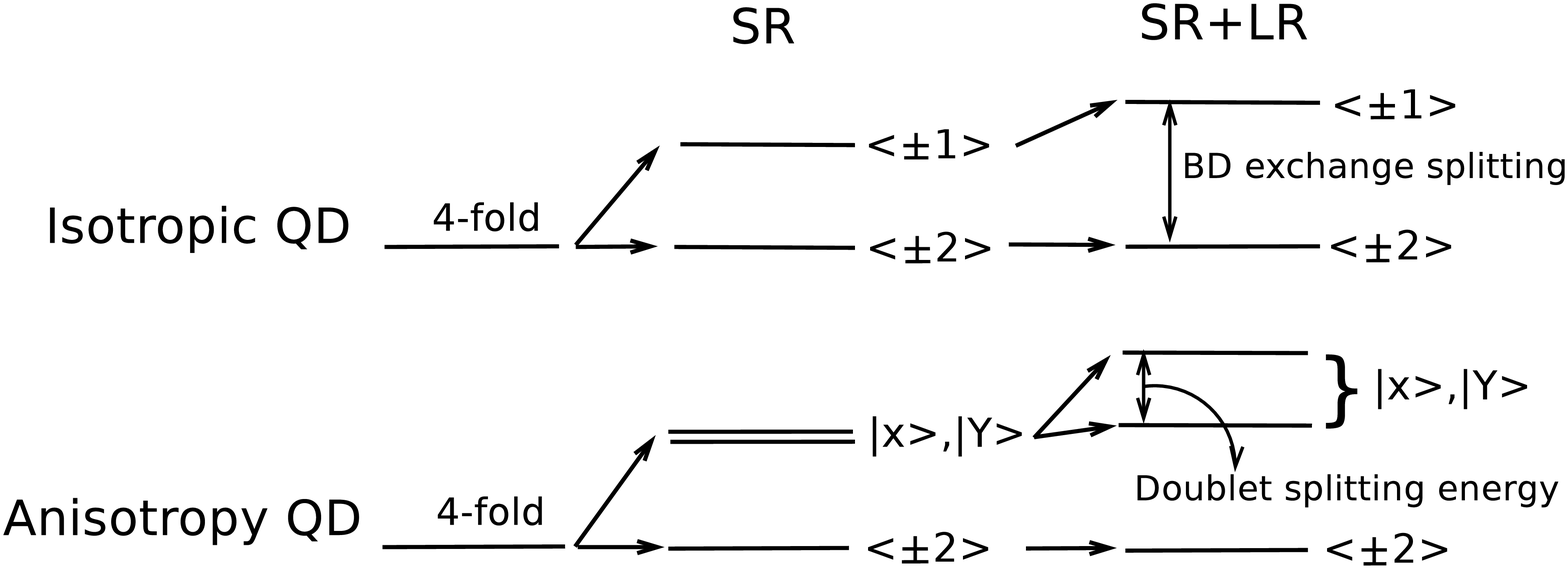}
  \end{center}
  \caption{The exciton fine structures in isotropic and anisotropic QDs are
    shown schematically for the cases with no e-h exchange interaction (left
    column), with only the SR exchange interaction (middle column) and with both
    the LR and SR exchange interactions (right column). Note that the doublet splitting
    energy with only the SR exchange interaction is very small and
    has been exaggerated in the figure.}
  \label{figtw1}
\end{figure}

In this section, we investigate the exciton fine structure in single
GaAs, InAs and GaN QDs. The fourfold degenerate
exciton ground states are split into a $J_z=\pm 2$ dark doublet and two bright
states when the e-h exchange interaction is taken into account. In a circular
QD, the two bright states are degenerate with the $z$-component of total angular
momentum $J_z=\pm 1$. For an anisotropic QD, the e-h exchange interaction
couples these two states together, forming the so-called $|X\rangle$ and
$|Y\rangle$ exciton states. 
These exciton fine structures are schematically shown in
  Fig.~\ref{figtw1}. The dark exciton doublet does not split since the e-h
exchange interaction is absent for  $J_z=\pm 2$ states
  [Eqs.~(\ref{equ:Ham_long}) and (\ref{equ:Ham_ss})].\cite{note8}
It is noted that in our results the exciton 
ground state is always the dark doublet,
which  is  consistent with the common
understanding.\cite{Bonneville,Ivchenko5,Bayer1} We restrict ourselves in
  discussing the fine structures split from the originally fourfold-degenerate
exciton ground states when the e-h exchange interaction is introduced, unless
otherwise specified.

\subsection{GaAs and InAs}
We first study the exciton fine structures in GaAs and InAs QDs. The doublet
splitting energy and the BD exchange 
splitting are calculated and scaling analyses are performed. 

\subsubsection{Doublet splitting}
\begin{figure}[htb]
  \begin{center}
    \includegraphics[width=7cm]{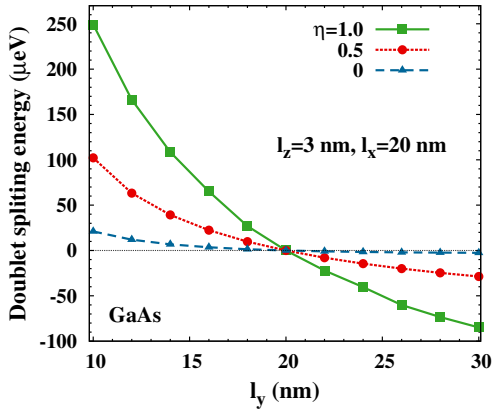}
  \end{center}
  \caption{(Color online) The doublet splitting energies in
    single GaAs QD as  function of the dot major/minor diameter
   $l_y$ with fixed  minor/major diameter $l_x=20$~nm and dot height $l_z=3$~nm.
 Different values of $\eta$
    are chosen to modulate the strength of the direct Coulomb interaction.} 
  \label{figtw2}
\end{figure}

In order to investigate the influence of the direct Coulomb
interaction on the doublet splitting energy, we introduce a dimensionless
parameter $\eta$ in front of the direct Coulomb interaction $U^{\rm eh}$ in
Eq.~(\ref{equ:Ham_total}). By varying $\eta$ from 0 to 1, the 
direct Coulomb interaction is varied. 
In the calculation, the major/minor diameter along the $x$-direction
$l_x$ is fixed at 20~nm and the minor/major diameter along the $y$-direction $l_y$ is varied from 
10 to 30~nm. The disk-like dot height $l_z$ is fixed at 3~nm. In our model, the
ground state of bright exciton is found to polarize along the axis of the weaker
confinement, i.e., $|X\rangle$ state when $l_x > l_y$ and $|Y\rangle$
state when $l_x<l_y$. Here the $|X\rangle$ 
and $|Y\rangle$ states are defined by their main components. The doublet splitting energy is
defined as $E_Y-E_X$ here and hereafter, with $E_X$ and
$E_Y$ representing the eigenenergies of $|X\rangle$ and
$|Y\rangle$ exciton states, respectively.
From Fig.~\ref{figtw2}, one observes that the doublet splitting is markedly
reduced with the decrease of the direct Coulomb interaction. When the 
direct Coulomb interaction is 
totally switched off, the doublet splitting energy is only less than $10\%$ of
its original value where the direct Coulomb interaction is fully taken care
of. The strong dependence of doublet splitting  
energy on the strength of direct Coulomb interaction is explained as follows. The 
direct Coulomb interaction attracts the electron and hole together and 
enhances the overlap of their wave functions. Hence according to
Eqs.~(\ref{equ:Ham_long23}) and (\ref{equ:Ham_short23}), with stronger direct Coulomb
interaction, stronger e-h exchange interaction is obtained. 

Another important feature in Fig.~\ref{figtw2} is that the doublet splitting
energy strongly depends on the dot shape.  
The absolute value of the doublet splitting energy decreases with decreasing dot
anisotropy, and tends to zero when the confining potential approaches
isotropic. It is seen from the curve with $\eta=1$ that the doublet splitting
energy varies from 0 to about 250~$\mu$eV when $l_y$ is varied from
20~nm(=$l_x$) to 10~nm. 

Our results are much larger than those reported very recently by
Kadantsev and Hawrylak,\cite{Kada} where they calculated the doublet splitting
energy with exciton wave function constructed by the product of 
single-particle ground-state wave functions of electron and hole 
states. So the effect of the direct Coulomb interaction was
totally absent and the doublet splitting was hence underestimated. Detailed comparison with the 
results in Ref.~\onlinecite{Kada} is given in Appendix~D. 

Our results are in good agreement with the existing experimental
  results.\cite{Gammon,Abbarchi,Bayer2} For example, 
the doublet splitting energy measured by Gammon {\em et al}.\cite{Gammon} in 
single GaAs QD lies in the range 20$\sim$50~$\mu$eV, which corresponds
approximately to $l_y=16$$\sim$18~nm, $l_x=20$~nm and $l_z=3$~nm in our
model as shown in Fig.~\ref{figtw2}. Good agreement is also reached with 
former  theoretical work by Takagahara\cite{Taka2} based on 
the variational method with the direct Coulomb 
interaction included, which also showed good agreement with the same
experiment.\cite{note4} 

\begin{center}
\begin{table}[htb]
\caption{Size parameters used in the calculation of doublet splitting energy in
  single InAs QD for Fig.~\ref{figtw3}(b).}
\begin{tabular}{ccccccccc}
\hline
\hline
     &\ \ \ 1\ \ \ &\ \ \ 2\ \ \ &\ \ \ 3\ \ \ &\ \ \ 4\ \ \ &\ \ \ 5\ \ \ &\ \ \ 6\ \ \ &\ \ \ 7\ \ \ \\
$l_x$ (nm)&21&19&17&15&14&13&12\\
$l_z$ (nm)&5.5&5.4&5.3&5.2&5.1&5&4.9\\
\hline
 &8&9&10&11&12&13&\\
$l_x$ (nm)&11.5&11&10.5&10&9.5&9\\
$l_z$ (nm)&4.8&4.75&4.7&4.55&4.5&4.4\\
\hline
\hline
\end{tabular}
\label{table2}
\end{table}
\end{center}

\begin{figure}[htb]
  \begin{minipage}[]{10cm}
    \hspace{-1.5 cm}\parbox[t]{5cm}{
      \includegraphics[width=4.5cm,height=4.5 cm]{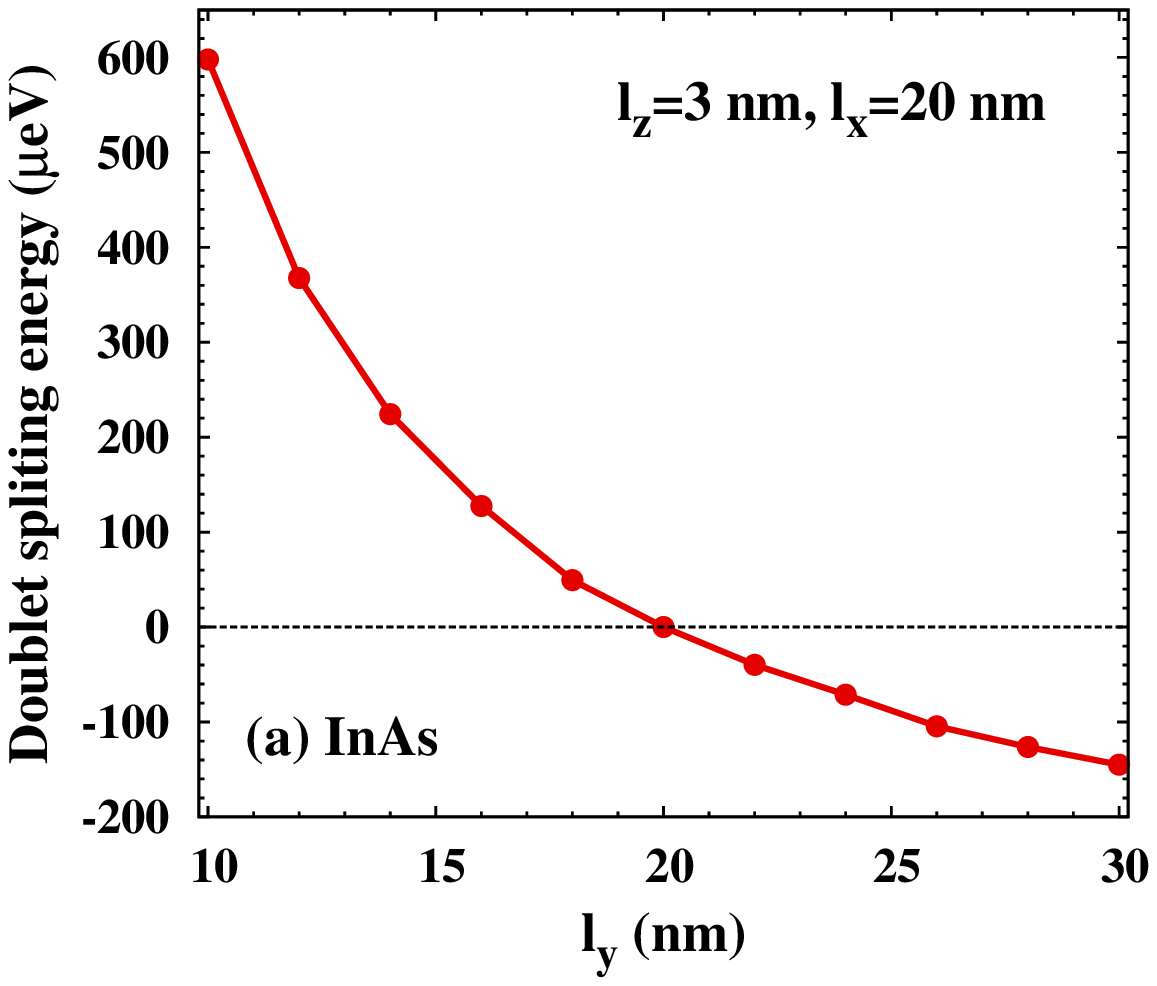}}
    \hspace{-0.7 cm}\parbox[t]{5cm}{
      \includegraphics[width=4.5cm,height=4.5 cm]{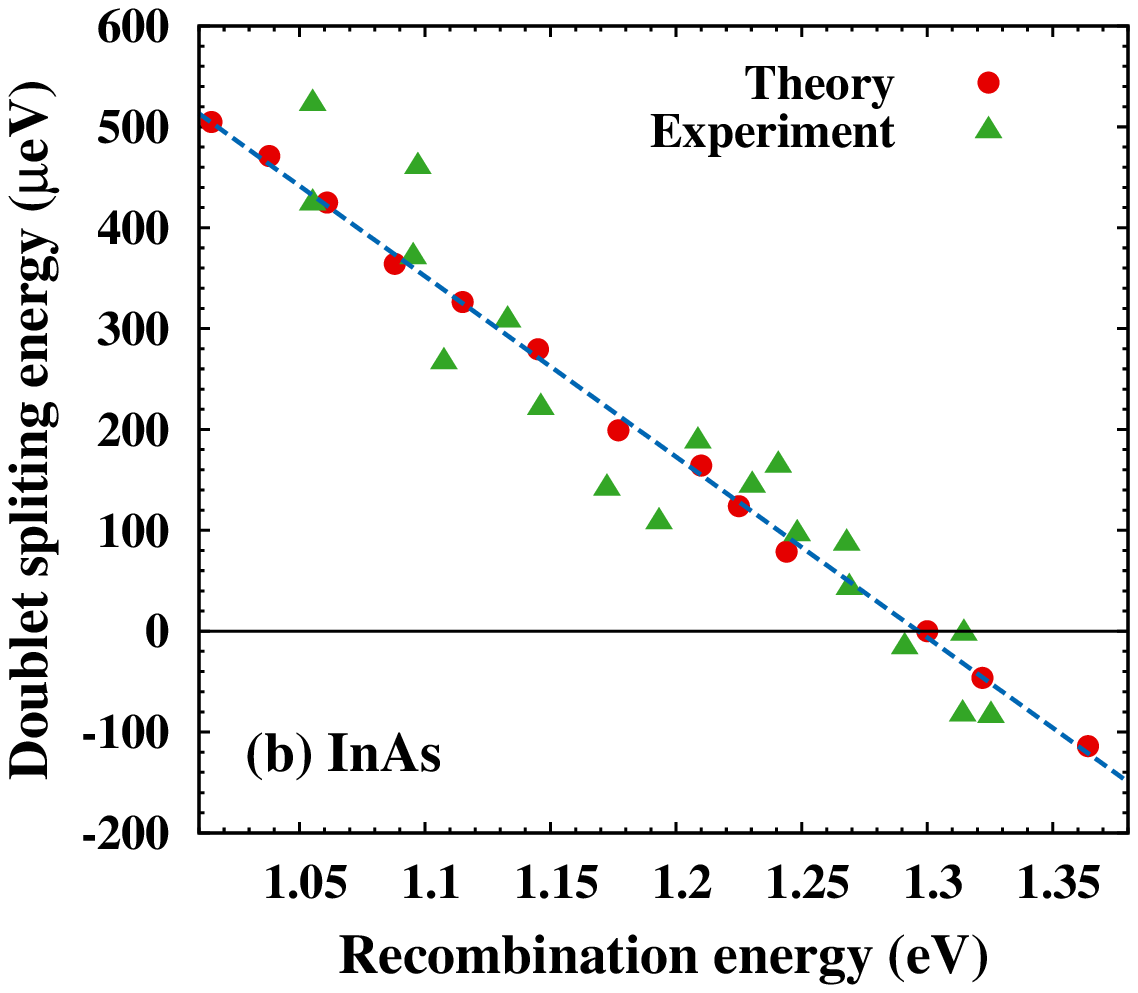}}
  \end{minipage}
  \caption{(Color online) The doublet splitting energy in single
    InAs QD (a) as a function of the dot major/minor diameter $l_y$
  with fixed minor/major diameter $l_x=20$~nm and dot height
    $l_z=3$~nm; (b) as a function of the exciton recombination
    energy. The experimental points are taken from 
    Ref.~\onlinecite{Seguin} and the theoretical points are calculated with
 the size parameters listed in Table~II. 
The dotted line is plotted for the guide of
    eye.}
  \label{figtw3}
\end{figure}

Similar features of the shape dependence of doublet splitting energy
are shown in Fig.~\ref{figtw3}(a) for InAs QDs with the direct Coulomb
  interaction fully included. In addition, we 
plot the doublet splitting energy as a function of the exciton recombination
energy in Fig.~\ref{figtw3}(b) in order to compare with the experimental data by
Seguin {\em et al}.\cite{Seguin}. The exciton recombination energy is defined as
$\frac{E_X+E_Y}{2}+E_g$, with $E_g$ standing for the band gap in cubic InAs. The
experimental data are 
taken from Ref.~\onlinecite{Seguin}. 
In the calculation, we fix 
the major/minor diameter $l_y=10$~nm and vary the minor/major diameter $l_x$ in
the range of $8.5\sim 20$~nm. The dot height $l_z$ is varied in the range of 
$4.3\sim 5.5$~nm. The exciton recombination energy is mainly modulated by
the strong confinement along the $z$-direction while the doublet splitting
energy is mainly determined by 
the ratio of $l_x:l_y$. So the
theoretical points are obtained by properly choosing the
 values of ($l_x$, $l_z$)
in pairs. From left to right  in Fig.~\ref{figtw3}(b), the 
explicit values of ($l_x$, $l_z$) for the theoretical points  
are given in order in Table~II.
It is seen from the figure that our theoretical results are in good agreement with the
experimental data. The doublet splitting energy decreases with increasing exciton
recombination energy and intersects through zero.  
Generally speaking, we employ larger values of $l_x$
together with larger $l_z$, and hence smaller recombination energy. The zero
point of doublet splitting energy is 
reached at $l_x=10$~nm and $l_z=4.55$~nm. 
In this way, the trend of the variation of the doublet splitting energy
with  the exciton recombination
energy is well explained as a result of the dot geometry.\cite{note9} 

The relative importance of  the LR and SR exchange interactions with respect
to the exciton fine structure splittings in QDs is an open question. Early
  work by Efros {\em et al.} included only the SR exchange interaction to
 investigate the band-edge excitons in spherical QDs. Takagahara\cite{Taka2}
assigned the origin of exciton doublet structure to  
the LR exchange interaction and Glazov {\em et
  al.}\cite{Ivchenko3} took into account only the LR exchange interaction in their
work to investigate exciton fine structure in an anisotropic QD.
However, Tsitsishvili {\em et al.}\cite{Kalt1,Kalt2} and Horodysk\'a {\em et
  al.} in their very recent work\cite{Horo} took into account only the SR
exchange interaction for excitons in anisotropic and spherical QDs, 
respectively. Here we reexamine the 
relative importance of the LR and SR exchange interactions
to the doublet splitting energy. Those to the BD exchange splitting are to be
discussed in the following corresponding parts.

\begin{figure}[htb]
  \begin{center}
    \includegraphics[width=7cm]{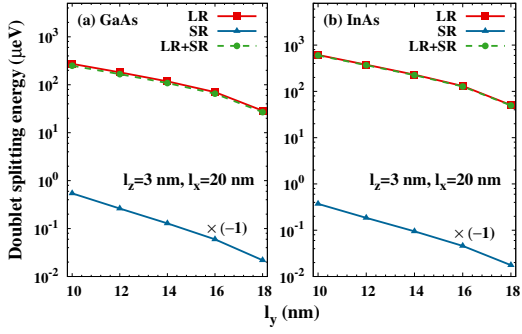}
  \end{center}
  \caption{(Color online) The doublet splitting energies from the LR, SR and
    both LR and SR exchange interactions, are plotted 
    as a function of the dot major/minor diameter $l_y$, with fixed
      minor/major diameter $l_x=20$~nm 
and dot height $l_z=3$~nm for (a) GaAs QD and (b) InAs QD.} 
  \label{figtw4}
\end{figure}

We calculate the doublet splitting energies by including the LR, SR and
  both LR and SR exchange interactions. In
Fig.~\ref{figtw4}, we plot the doublet splittings as a function of the
major/minor diameter $l_y$ with the minor/major diameter $l_x$ fixed in GaAs and
InAs QDs. It is seen that the curves
obtained by including the LR exchange interaction only and by including
both the LR and SR exchange interactions almost match each other. The
doublet splitting energies from the SR exchange interaction
are always less than $1~\mu$eV, more than two orders of magnitude 
smaller than the 
splittings caused by the LR exchange interaction. So the LR exchange interaction
is dominant in determining the doublet splitting energy when all heavy-hole,
light-hole and split-off bands are taken into account. This is
consistent with Ref.~\onlinecite{Taka2} where only the heavy- and light-hole bands
are taken account of.

It is also noted that the doublet splitting energies from the SR
exchange interaction are in opposite sign to those from the LR exchange
interaction. This means that the relative positions of
$|X\rangle$ and $|Y\rangle$ exciton states are reversed when only the SR
exchange interaction is included. 

\begin{figure}[htb]
  \begin{minipage}[]{10cm}
    \hspace{-1.5 cm}\parbox[t]{5cm}{
      \includegraphics[width=4.5cm,height=4.5 cm]{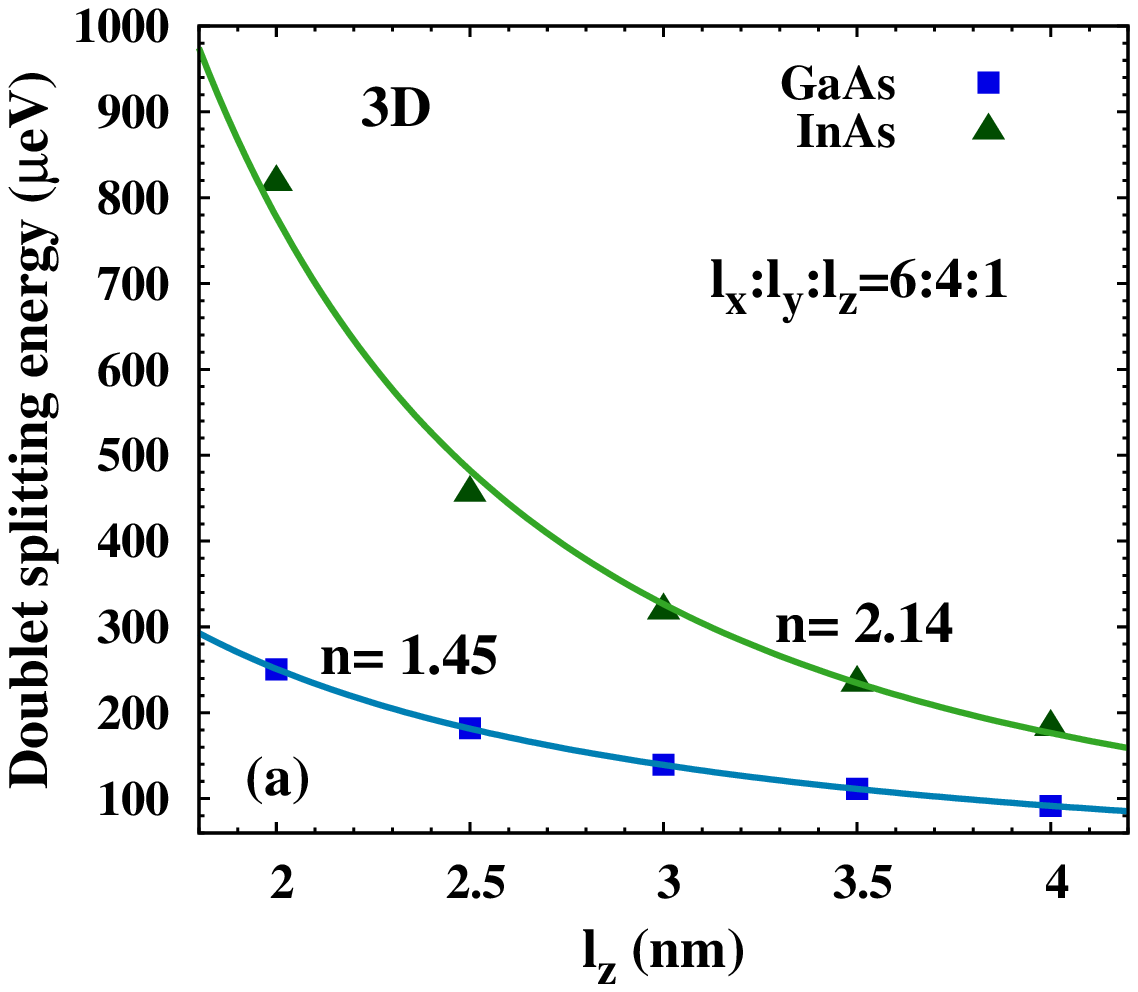}}
    \hspace{-0.7 cm}\parbox[t]{5cm}{
      \includegraphics[width=4.5cm,height=4.5 cm]{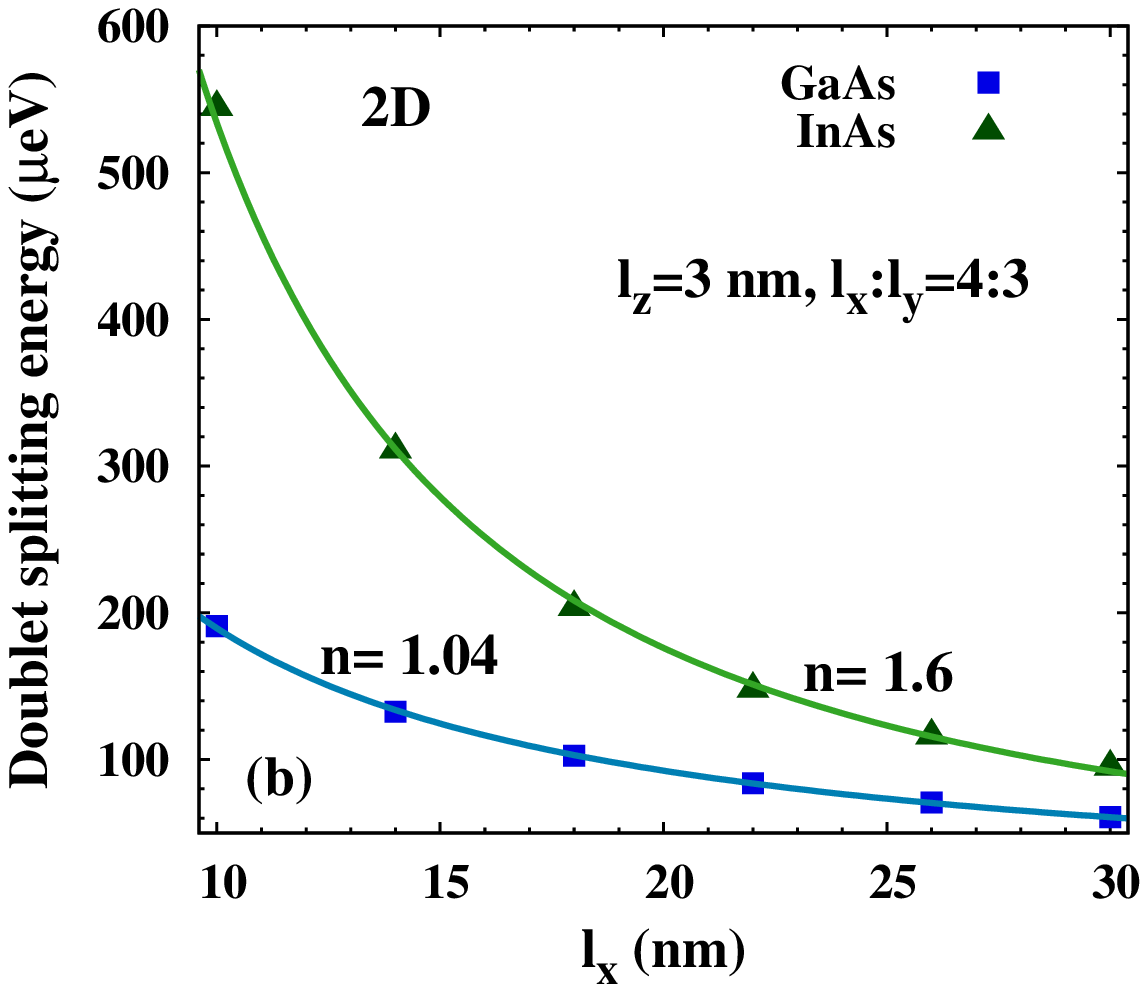}}
  \end{minipage}
  \caption{(Color online) The doublet splitting energies in GaAs and
    InAs QDs. (a) For the 3D scaling, with $l_x:l_y:l_z=6:4:1$. 
 (b) For the 2D scaling, with $l_x:l_y=4:3$ and  $l_z=3$~nm. The
    solid curves are fit to the power law $C/L^n$ and the index 
    values $n$ are shown next to the curves. }
  \label{figtw5}
\end{figure}

We further investigate the size dependence of the doublet splittings 
of  GaAs and InAs QDs and perform the 3D
and 2D size-scaling analysis. Detailed results
are plotted in Fig.~\ref{figtw5} with the solid curves  fit to the power 
law $\propto 1/L^n$. The results
are consistent with the physical intuition. 
The doublet splitting energy increases with
the decrease of the dot size since for smaller dot size, 
the overlap of the wave functions
of electron and hole is enhanced and hence 
larger matrix elements of e-h exchange
interaction are obtained [Eqs.~(\ref{equ:Ham_long23}) 
and (\ref{equ:Ham_short23}) and their
Hermitian conjugates]. We obtain $n_{\rm GaAs}=1.45$ and $n_{\rm InAs}=2.14$
in the 3D scaling 
and $n_{\rm GaAs}=1.04$ and $n_{\rm InAs}=1.6$
in the 2D scaling. All power indices
obtained lie in the range of 1$\sim$3 and are consistent with the scaling rules
for  the LR exchange interaction established in Sec.~II. 
The fact that the power indices of InAs QDs are larger than those of GaAs QDs is
understood as follows. The exciton Bohr radius is 14.9~nm in bulk GaAs and
51.6~nm in bulk InAs. As compared to the characteristic size $L$ of  
the QDs in the 3D and 2D scalings, for GaAs QDs,
$a_{\rm Bohr}^{\rm GaAs}$ is comparable to or smaller than $L$. So GaAs QDs
are closer to the weak confinement limit.
 For InAs QDs, $a_{\rm Bohr}^{\rm InAs}$ is larger than $a_{\rm Bohr}^{\rm GaAs}$,
 so InAs QD is  closer to the strong confinement limit compared to GaAs QD.

\subsubsection{BD exchange splitting}
The band-edge exciton states in isotropic QDs were investigated in previous
works\cite{Efros,Horo} by including only the SR exchange interaction. 
It is noted that even though the doublet splitting energy, which is mainly from
the LR exchange interaction, becomes zero in
circular QDs, the LR exchange interaction still contributes to the splitting between the bright
and dark exciton states. 
For circular GaAs QD with $l_z=3$~nm and
$l_x=l_y=20$~nm, the BD exchange splitting is 1.76~meV and the LR exchange interaction contributes
0.17~meV to it. So in GaAs QDs, the SR exchange interaction is dominant in
determining the BD exchange splitting. It is a 
different case in InAs QD of the same size, where the LR exchange interaction
contributes 0.31~meV out 
of the total BD exchange splitting 0.49~meV. This difference is due
to the fact that the singlet-triplet splitting parameter $\Delta E_{\rm SR}$ is
20~$\mu$eV in GaAs compared 0.3~$\mu$eV in
InAs. As a result, the SR exchange interaction is rather weak in InAs. The
calculated BD exchange splitting of 500~$\mu$eV in InAs QD with $l_z=6$~nm and
$l_x=l_y=15$~nm is in good agreement with the experimental data.\cite{Bayer1} So we
emphasize that the LR exchange interaction is important 
to understand the experimental results not only for the doublet splitting energy, but
also for the BD exchange splitting in QDs.

\begin{table}
\caption{ Power indices of the 2D and 3D scalings of the BD exchange splitting
  contributed by the LR, SR and both LR and SR
  exchange interactions in GaAs and InAs 
  QDs.} 
\begin{tabular}{c|ccc|ccc}
\hline\hline
          &         &  3D   &      &    & 2D  &\\
    \hline
      &\ \  LR\ \  & \ \ \  SR\ \  \ & \  Both\ \  &\ \  LR\ \   &\ \  \ SR  \ \ \ & \ Both\ \ \\
GaAs  &     1.57       &  1.52   &   1.49      &   1.14    &   0.15   & 0.22  \\      
InAs  &      1.95      &   2.00       &  1.95     &   1.69    & 0.80 &1.38\\
\hline\hline
\end{tabular}
\label{table3}
\end{table}

\begin{figure}[htb]
  \begin{minipage}[]{10cm}
    \hspace{-1.5 cm}\parbox[t]{5cm}{
      \includegraphics[width=4.5cm,height=4.5 cm]{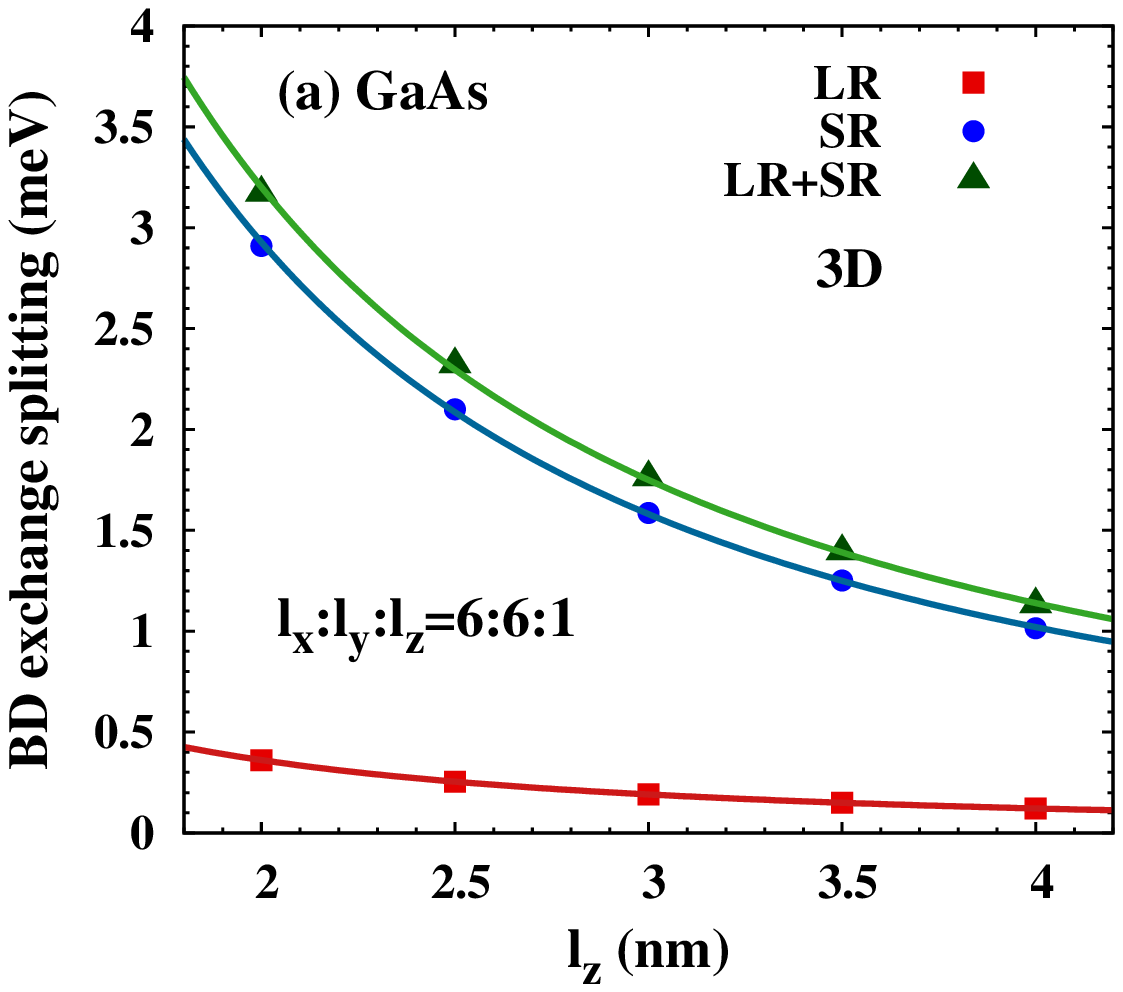}}
    \hspace{-0.7 cm}\parbox[t]{5cm}{
      \includegraphics[width=4.5cm,height=4.5 cm]{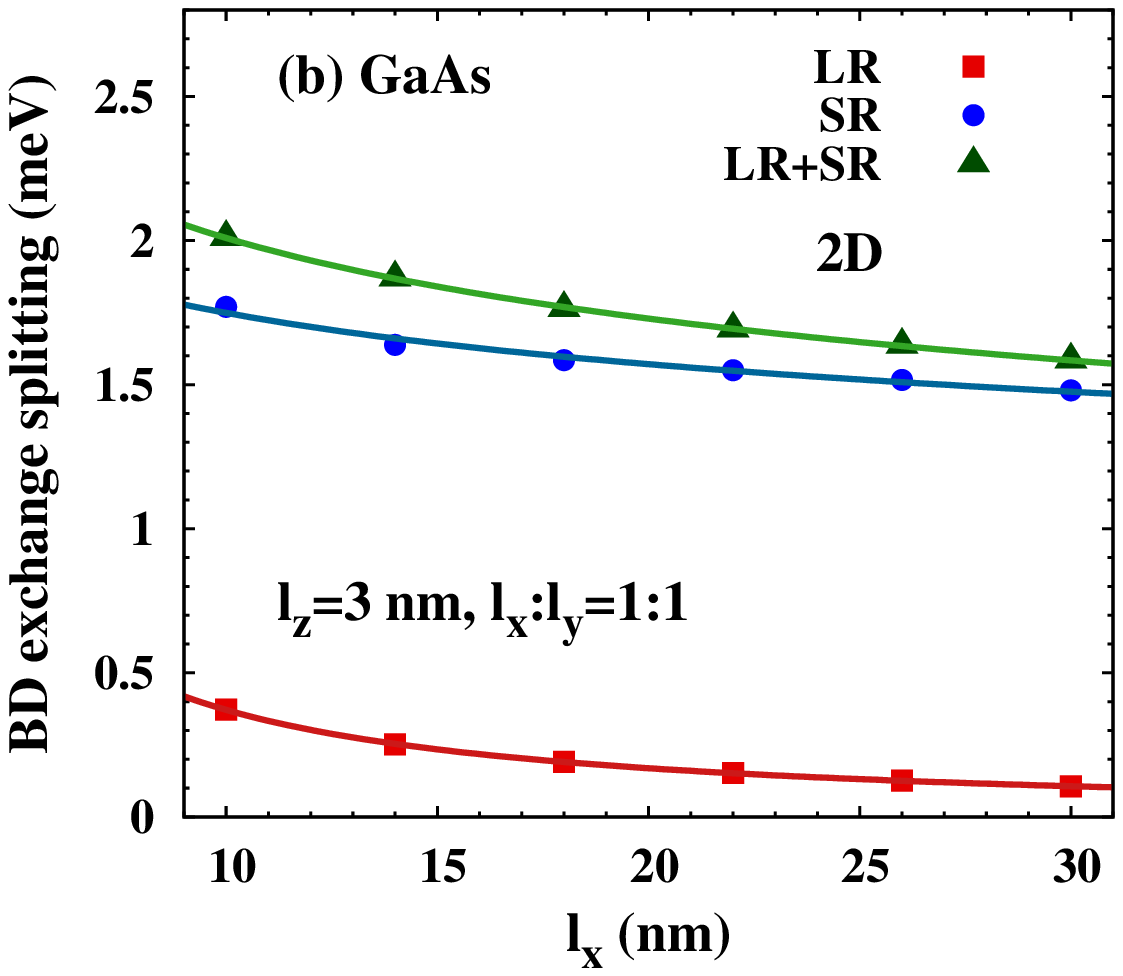}}
  \end{minipage}
  \begin{minipage}[]{10cm}
    \hspace{-1.5 cm}\parbox[t]{5cm}{
      \includegraphics[width=4.5cm,height=4.5 cm]{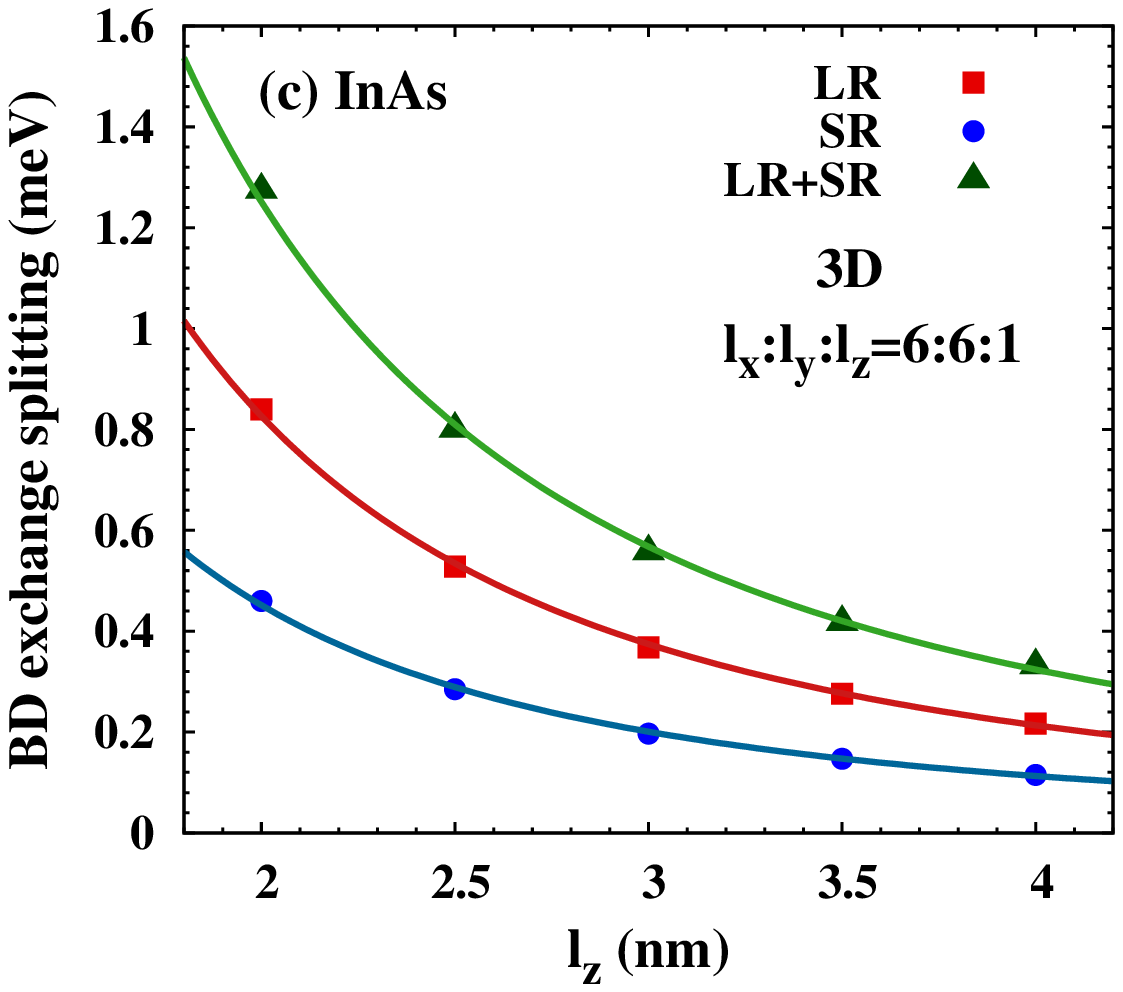}}
    \hspace{-0.7 cm}\parbox[t]{5cm}{
      \includegraphics[width=4.5cm,height=4.5 cm]{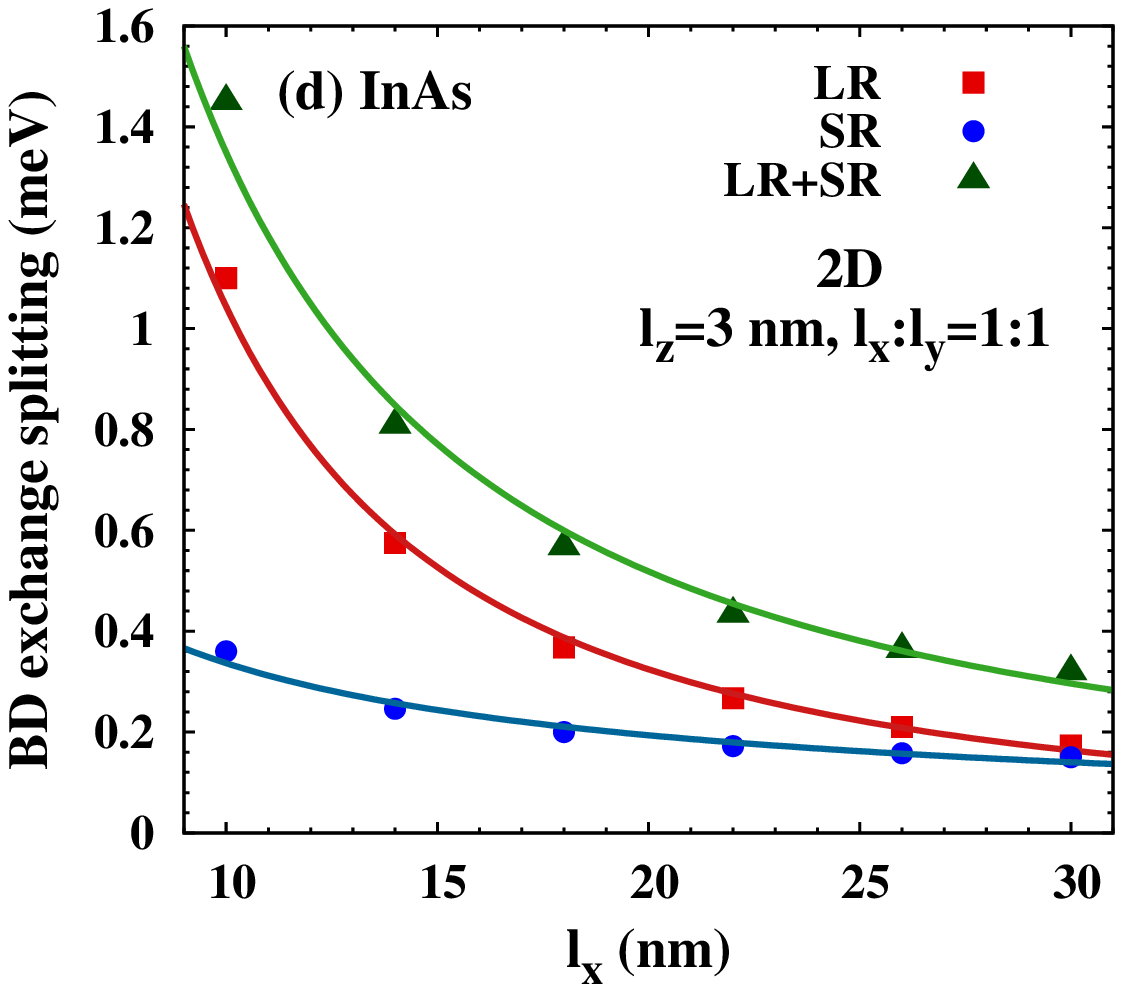}}
  \end{minipage}
  \caption{(Color online) The BD exchange splittings in circular QDs by taking
    account of the LR, SR  and both LR and SR exchange interactions. (a)
    For the 3D scaling in GaAs QD, with $l_x:l_y:l_z=6:6:1$. (b) For the 2D
    scaling in GaAs QD with $l_x=l_y$ and
      $l_z=3$~nm. (c) For the 3D scaling in InAs QD, with
  $l_x:l_y:l_z=6:6:1$. (d) For the 2D  scaling in InAs QD,
    with $l_x=l_y$ and  $l_z=3$~nm. The solid curves are fit to the power law
    $C/L^n$ with $n$ listed in Table III.} 
  \label{figtw6}
\end{figure}

The size dependence of the BD exchange splitting is investigated in circular QDs for the purpose of
eliminating the effect of doublet splitting. In the calculation, we take into
account the LR, SR and both LR and SR exchange interactions.
The results are shown in Fig.~\ref{figtw6}. The solid curves are fit to the power 
law $\propto 1/L^n$, and the obtained power indices of the size scaling of BD exchange 
splitting are listed  in Table~III. 
The size scaling laws of the BD exchange splitting by including only the LR
exchange interaction are close to those
of the doublet splitting energy in both the 3D and 2D scalings in GaAs and InAs
QDs (see Fig.~\ref{figtw6} and Table III). This is because the diagonal
and off-diagonal matrix elements of the LR 
exchange interaction in Eq.~(\ref{equ:Ham_long}) are in similar forms of the
${\bf r}$ dependence. The power indices of  
the BD exchange splitting from the SR exchange interaction lie reasonably in the
range 1$\sim$3 in the 3D scaling and in the range 0$\sim$2 in the 2D scaling, and
are consistent with the scaling rules established in Sec.~II. 

\begin{figure}[htb]
  \begin{center}
    \includegraphics[width=7.5cm]{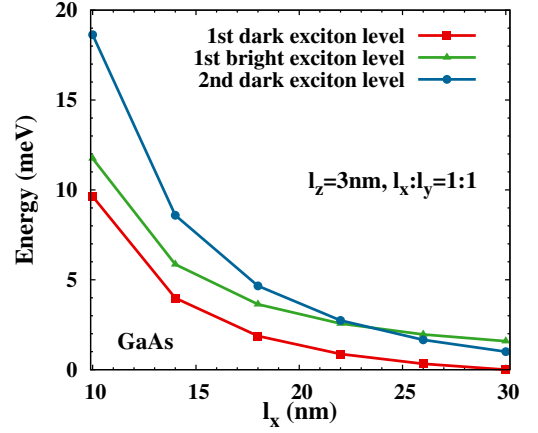}
  \end{center}
  \caption{(Color online) The energy positions of 
the lowest dark and bright levels
    and the first excited dark level as function of $l_x$, with 
$l_x=l_y$ and $l_z=3$~nm. The zero point of energy is set at the 
    lowest energy level at $l_x=30$~nm.}
  \label{figtw7}
\end{figure}

An important feature of the relative energy positions of the dark and bright exciton
states can be deduced from the above scalings. As in the 2D
scalings, the power indices of the BD exchange splitting in isotropic GaAs and InAs QDs
(also in GaN QD, see in the next subsection) are smaller than 2 in the size
range under investigation. Meanwhile the
level spacing between ground dark level and first excited dark level, which is
induced by the lateral confinement, scales approximately as $\propto
\frac{1}{L^2}$ [Eq.~(\ref{equ:W-xy})].\cite{note5} So as the dot diameter increases, the
first excited dark exciton level decreases more rapidly than the ground bright exciton 
level, and a crossing between these two levels may occur. For a typical case, in Fig.~\ref{figtw7} we plot the
eigenenergies of the ground dark exciton and bright exciton levels 
and the first excited dark exciton level as a function of the dot diameter $l_x$(=$l_y$) in GaAs QD. The dot
height $l_z$ is fixed at 3~nm. When the dot diameter $l_x$ is varied from 10 to
30~nm, a crossing between the ground bright doublet and the first
excited dark quartet (since $l_x=l_y$, the first excited dark exciton level is
fourfold-degenerate) is observed around $l_x=23$~nm. 
This crossing can also be obtained in anisotropic QDs with the underlying physics. 

This size-dependent bright-dark exciton level crossing provides a unique way of tuning
the bright exciton states in resonance with the dark exciton states, which is
meaningful to recent research on the optical nuclear spin pumping with the
help of the hyperfine-interaction-mediated spin-flip transitions between the bright and dark exciton
states.\cite{Klotz,Chekhovich} 

\subsection{GaN}
We then investigate the properties of  the LR and SR exchange
interactions in cubic GaN QD where theoretical works are
absent. Compared to In$_{1-x}$Ga$_x$As-based  
structures, nanostructures based on GaN are less investigated 
both experimentally and theoretically. Much attention has been attracted for their
unique properties, e.g., the wide band gap ($\sim$3.3~eV), which represents great
potential for applications in electronics and photonics at 
temperature much higher than the liquid-helium or liquid-nitrogen cryogenic
temperatures.\cite{Bardoux,Kako,Kindel} 

\subsubsection{Doublet splitting}
\begin{figure}[htb]
  \begin{minipage}[]{10cm}
    \hspace{-1.5 cm}\parbox[t]{5.cm}{
      \includegraphics[width=4.5cm,height=5 cm]{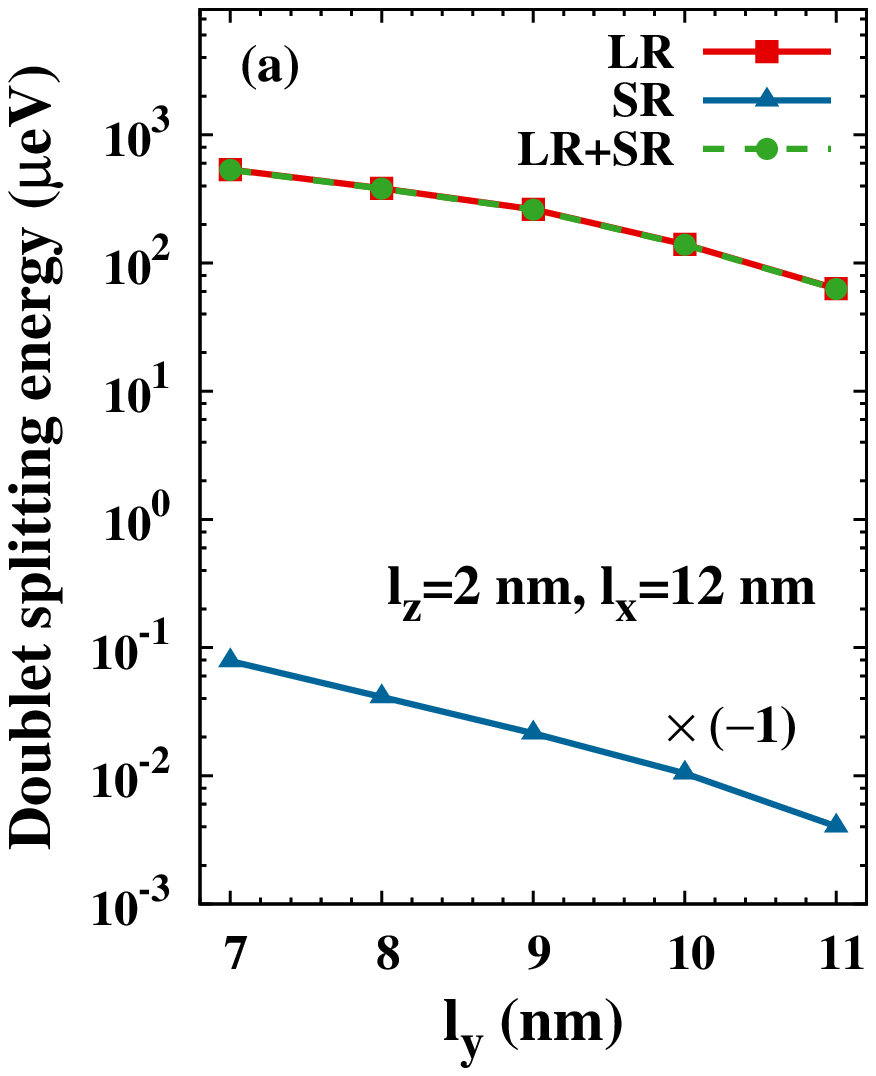}}
    \hspace{-0.7 cm}\parbox[t]{4.5cm}{
      \includegraphics[width=4.4cm,height=5 cm]{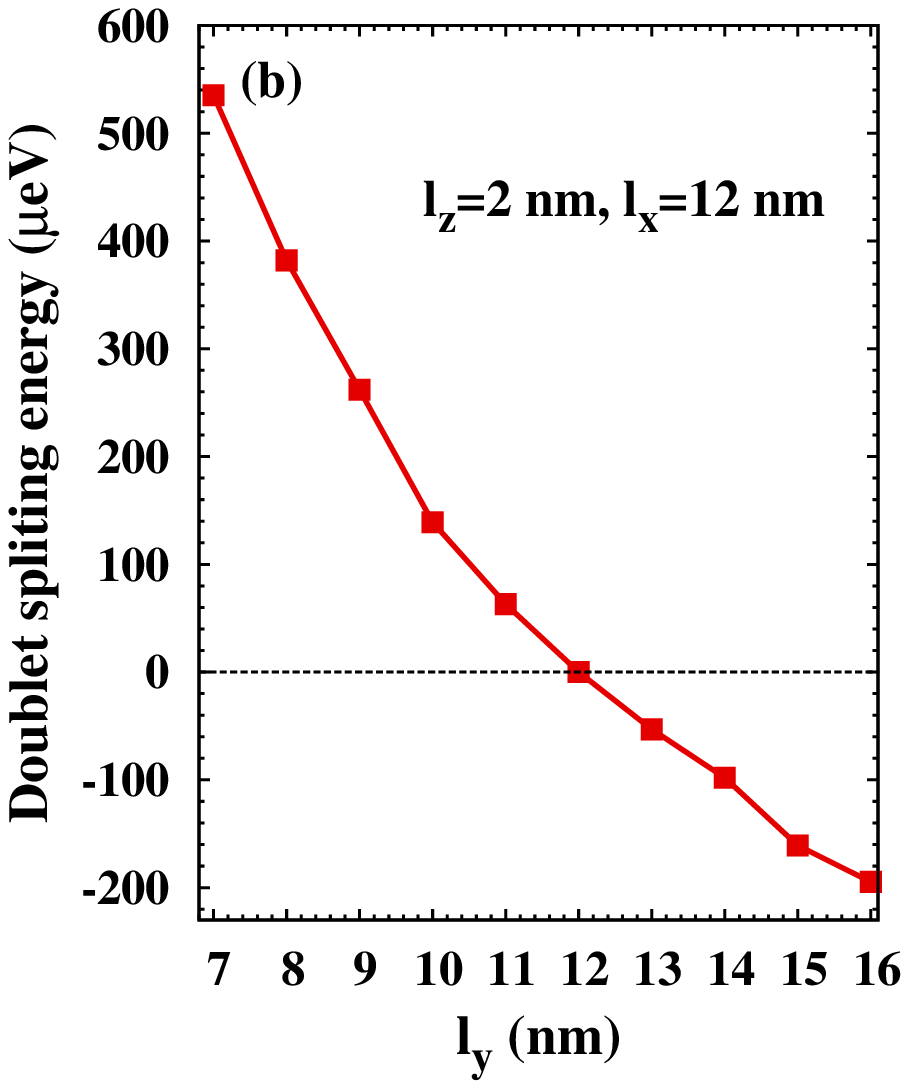}}
  \end{minipage}
  \caption{(Color online) The doublet splitting energies in single GaN QD
 as a function of the dot major/minor diameter $l_y$ with fixed
minor/major diameter $l_x=12$~nm and dot height $l_z=2$~nm. (a) The
      doublet splitting energies from the LR, SR exchange interactions and both
      together; (b) the
    doublet splitting energy from the LR exchange interaction.} 
  \label{figtw8}
\end{figure}

Due to the absence of experimental value of the singlet-triplet splitting
$\Delta E_{\rm SR}$ in cubic GaN, the doublet splitting energy in GaN QDs is
calculated by temporarily setting $\Delta E_{\rm SR}=20~\mu$eV.\cite{GaN-sr} 
In Fig.~\ref{figtw8}(a), we plot the doublet splitting energies calculated by
including the LR, SR and both LR and SR exchange interactions as a function of
the dot minor diameter $l_y$. Similar to GaAs and InAs
QDs, the SR exchange interaction is irrelevant when considering the
doublet splitting energy. Since the strength of the SR exchange interaction is
proportional to the value of $\Delta E_{\rm SR}$, we assert that $\Delta E_{\rm
  SR}$ in GaN can not be so large as to make 
the SR exchange interaction comparable or even exceed the LR exchange
interaction because otherwise it will make other results, 
e.g., the BD exchange splitting, unreasonable. So in Fig.~\ref{figtw8}(b), we plot
the doublet splitting energy varying with the dot shape, calculated
without the SR exchange interaction. 
Size parameters are chosen according to the experiment.\cite{Marie} 
One observes in the figure that with large anisotropy, the doublet splitting energy
reaches 100s of $\mu$eV. The large doublet splitting energy obtained is key
to understand the experiment by Lagarde {\em et al.}\cite{Marie} 
where the conversion from exciton linearly polarized
states $|X\rangle$ and $|Y\rangle$ to the circularly polarized ones $|\pm
1\rangle$ was not observed for magnetic field up to 4T.

\begin{figure}[htb]
  \begin{center}
    \includegraphics[width=7.5cm]{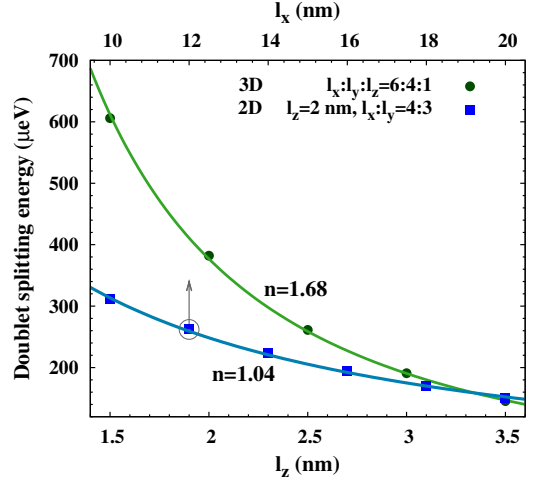}
  \end{center}
  \caption{(Color online) The doublet splitting energies in GaN
    QDs. 1) For 3D scaling, with $l_x:l_y:l_z=6:4:1$. $l_z$ is varied from
    1.5 to 3.5~nm. 2) For 2D scaling, with $l_x:l_y=4:3$ and $l_z=2$~nm. 
 The solid curves are fit to the power law $C/L^n$ with the index
 values $n$ shown next to the curves. Note the scale of the 2D scaling is on
top of the frame.}
  \label{figtw9}
\end{figure}

The size dependence of doublet splitting energy is also investigated. The
obtained results can be well 
understood in a straightforward way as those for GaAs and InAs QDs discussed above. So we only plot our
results in Fig.~\ref{figtw9} without more discussions. The power indices obtained
from the 2D and 3D size scalings are shown next to the curves.  The scaling laws of doublet
splitting energies found in GaN QDs are consistent with the rules established in Sec.~II.

\begin{figure}[htb]
  \begin{center}
    \includegraphics[width=8cm,height=7cm]{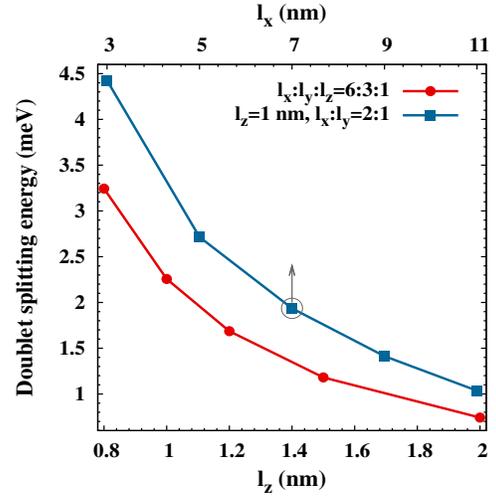}
  \end{center}
  \caption{(Color online) The doublet splitting energies in
    small cubic GaN QDs as 
function of dot size. Note the scale of one of the curves
    is on top of the frame.}  
  \label{figtw10}
\end{figure}

We further calculate the doublet splitting energy in small cubic GaN QDs by
pushing our model to its extreme.\cite{note6}  In Fig.~\ref{figtw10}, we plot the
doublet splitting energies as function of dot size. It is seen that the
doublet splitting energies reach several eV 
when the QDs become extremely small. This is strongly supported by recent
experiment on wurtzite GaN QDs where doublet splitting energies in the range of 
2$\sim$7 meV were reported.\cite{Kindel} Further experiments on cubic GaN QDs are
 expected.

\subsubsection{BD exchange splitting}
Although the SR exchange interaction is negligible concerning the doublet splitting in
GaN QDs, its contribution to the BD exchange
splitting is still comparable to that of the LR exchange interaction. 
In order to investigate the properties of BD exchange splitting in GaN QDs,
the singlet-triplet splitting $\Delta E_{\rm SR}$ is again set at $20~\mu$eV as a
parameter.\cite{GaN-sr} Due to the fact that in circular QDs,
the BD exchange splitting calculated by including both the LR and
SR exchange interactions is approximately the summation of those obtained by including
each separately and the contribution of the SR exchange interaction is proportional to the value of $\Delta
E_{\rm SR}$, the genuine value of $\Delta E_{\rm SR}$ in cubic GaN can be
extracted by comparing our
theoretical results of BD exchange splitting with future experimental data. 

\begin{table}
\caption{ Power indices of the 2D and 3D scaling of the BD exchange splitting
  contributed by  the LR,  SR and both LR and SR 
 exchange interactions in GaN QDs.} 
\begin{tabular}{c|ccc|ccc}
\hline\hline
          &         &  3D   &      &    & 2D  &\\
    \hline
      &\ \  LR\ \  & \ \ \  SR\ \  \ & \  Both\ \  &\ \  LR\ \   &\ \  \ SR  \ \ \ & \ Both\ \ \\
GaN   &     1.76      &    1.74       &  1.75   &   1.26     & 0.22  &0.68\\
\hline\hline
\end{tabular}
\label{table4}
\end{table}

\begin{figure}[htb]
  \begin{minipage}[]{10cm}
    \hspace{-1.5 cm}\parbox[t]{5cm}{
      \includegraphics[width=4.5cm,height=4.5 cm]{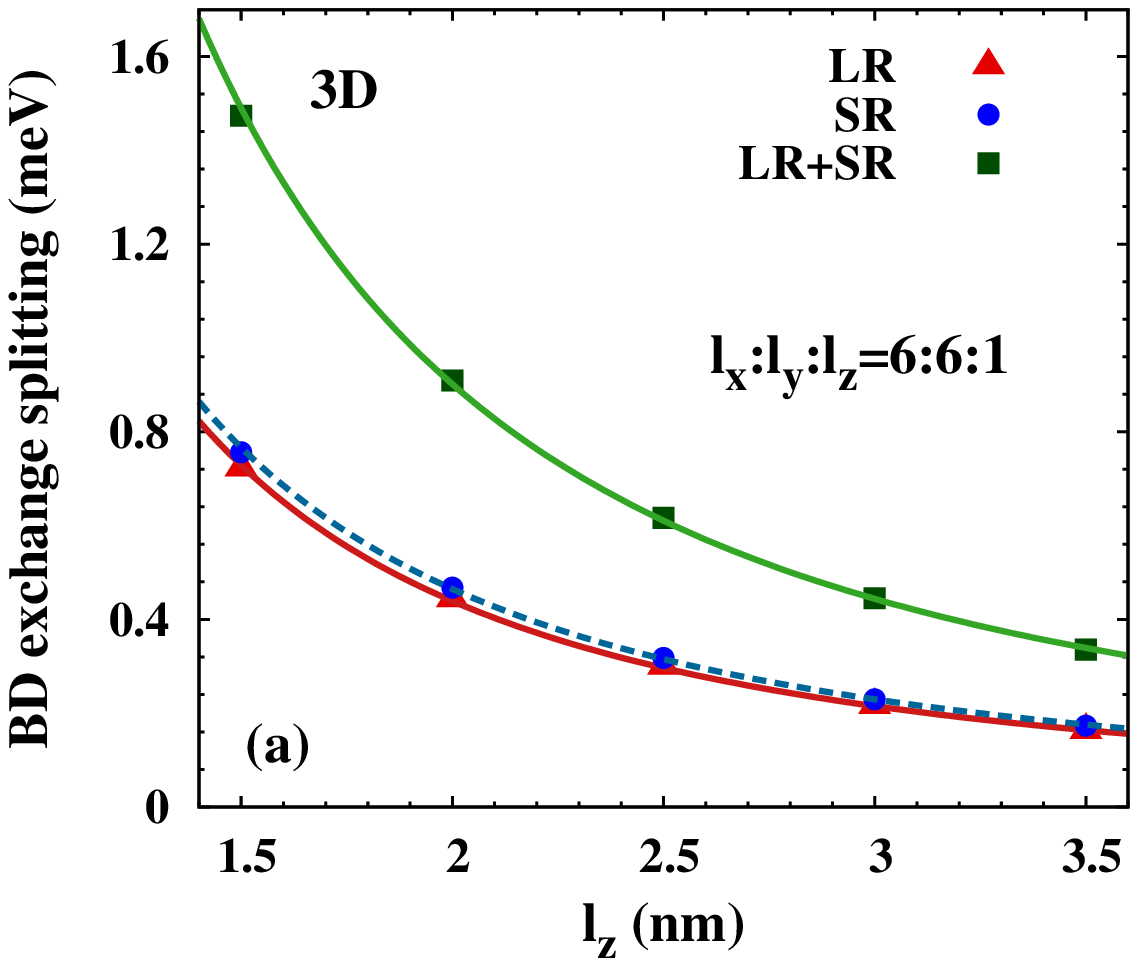}}
    \hspace{-0.7 cm}\parbox[t]{5cm}{
      \includegraphics[width=4.5cm,height=4.5 cm]{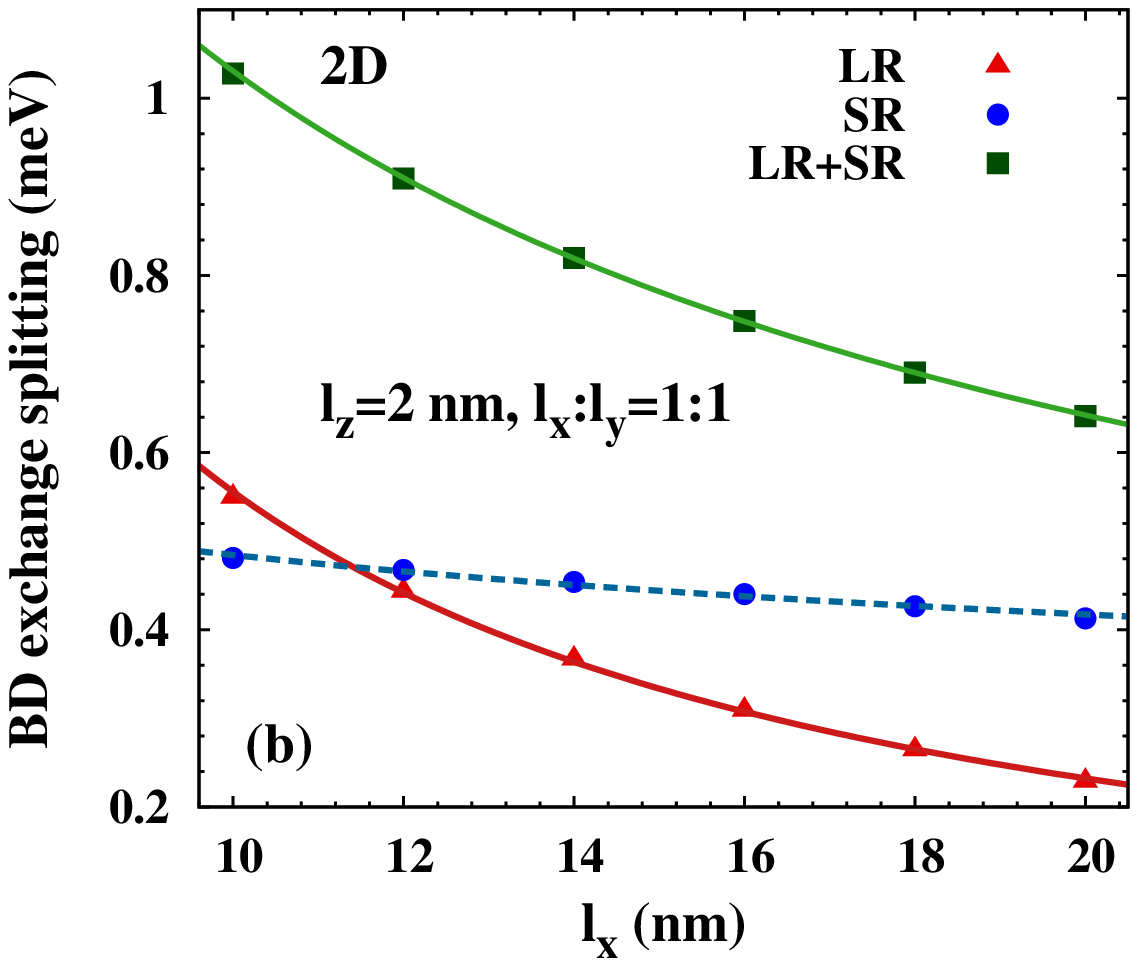}}
  \end{minipage}
  \caption{(Color online) The BD exchange splittings in GaN QDs by taking
    account of both  the LR and SR exchange interactions, and separately: (a)
    for the 3D
    scaling, with 
    $l_x:l_y:l_y=6:6:1$ and (b) for the 2D scaling, with 
   $l_x=l_y$ and $l_z=2$~nm. The solid and dashed curves are fit to
    the power law $C/L^n$ and the index values $n$ are shown next to the curves. }
  \label{figtw11}
\end{figure}

In Fig.~\ref{figtw11}, we plot the BD exchange splitting as function of the
dot size. The obtained power indices of size scaling are listed in Table IV. From
Fig.~\ref{figtw11}(b), one observes a crossing of BD exchange splittings
obtained from
the LR exchange interaction and from  the SR exchange interaction,  
indicating the exchange of  the relative importance of the LR and the SR exchange
interactions. This is due to the different scaling rules of the LR and
SR exchange interactions in the 2D 
scaling [Eqs.~(\ref{equ:scaling-L2}), (\ref{equ:scaling-L4}) and
(\ref{equ:scaling-S2}), (\ref{equ:scaling-S5})]. As the the dot diameter
increases, the BD exchange splitting from the LR exchange
interaction decreases more rapidly than that from the SR exchange interaction
(see in Table IV for detailed values of power indices) and becomes smaller than
it for $l_x>11.4$~nm in GaN QD. In fact, one may also expect a crossing 
in Fig.~\ref{figtw6}(d) in the region of $l_x>30$~nm in InAs QD. 

\subsection{Discussions on relative importance of the heavy-hole, light-hole and split-off bands}
\begin{figure}[htb]
  \begin{center}
    \includegraphics[width=8.5cm]{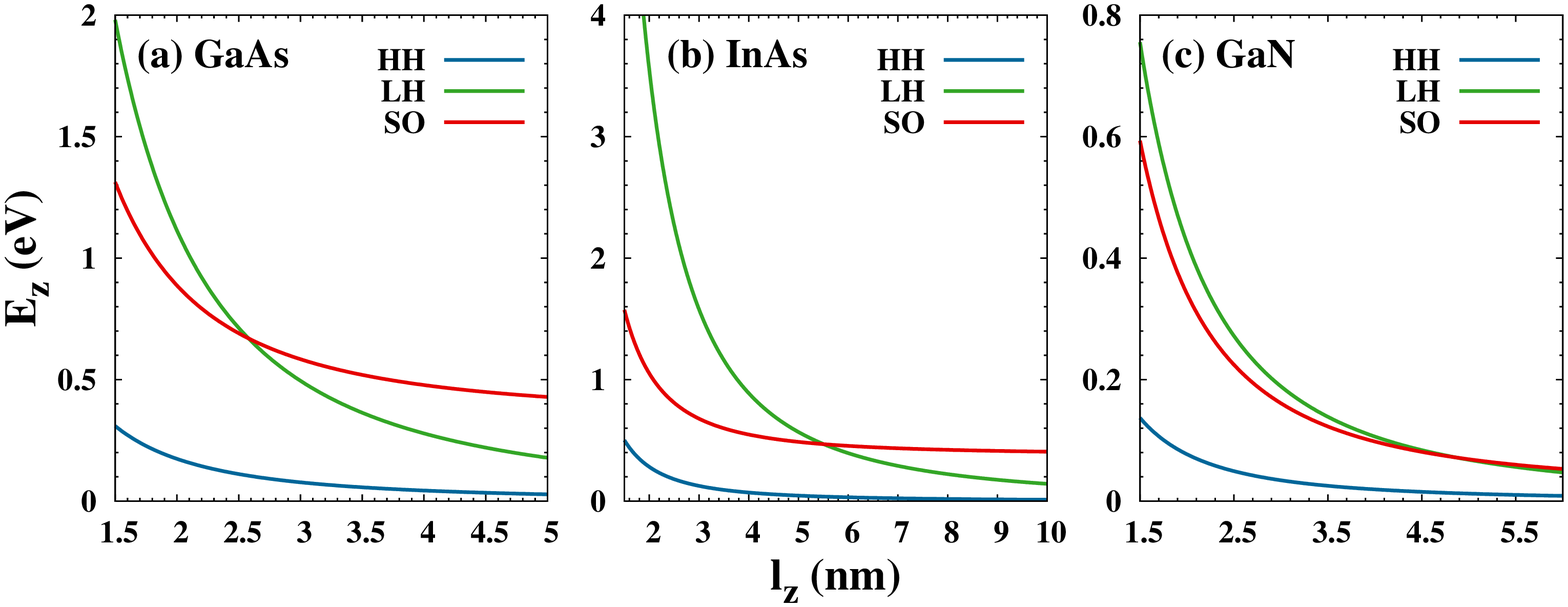}
  \end{center}
  \caption{(Color online) Eigenenergies of decoupled heavy-hole (HH),
    light-hole (LH) and
    split-off-hole (SO) ground states in QW. Infinite square well potential
    is employed and the well width is denoted as
    $l_z$. (a) GaAs, (b) InAs, (c) GaN. } 
  \label{figtw12}
\end{figure}

We now turn to address the relative importance of the heavy-hole, light-hole 
and split-off bands to the exciton fine structure.  
Theories of exciton fine structure in QDs were presented by taking into
account only the heavy-hole band.\cite{Andreani,Ivchenko4} It was pointed out by Takagahara that
it is important to take account of the mixing of heavy-hole and light-hole bands
to explain the experimental phenomena.\cite{Taka2} But this argument was not explicitly
proved in that paper. In Fig.~\ref{figtw12}, we plot the energies of the lowest
subbands of decoupled heavy, light and split-off holes in
GaAs, InAs and GaN QWs. The
  valence band coupling is 
  neglected so that $E_z=\frac{\hbar^2\pi^2}{2m_z^\ast l_z^2}$, with
$m_z^\ast$ representing the effective mass of the 
heavy, light or split-off hole in the
$z$-direction.
The relative positions of these three subbands vary with
$l_z$ because of different hole effective masses in the
$z$-direction.\cite{Winkler} As shown in the figure,
in the strong confinement regime, the split-off 
subband is closer to the heavy-hole subband than the light-hole one for all three
materials. Therefore, if the light-hole band is considered,
the split-off band should also be included.  

\begin{figure}[htb]
  \begin{center}
    \includegraphics[width=9cm]{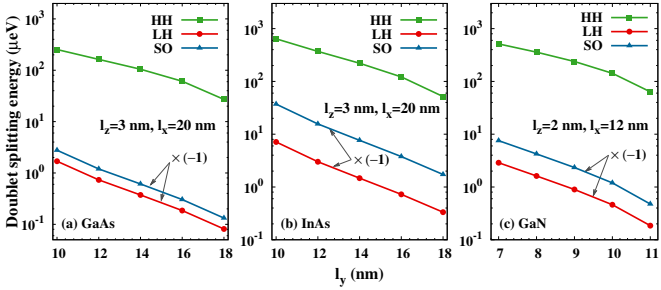}
  \end{center}
  \caption{(Color online) Contributions of the heavy-hole (HH), light-hole (LH) and
    split-off (SO) bands to the doublet splitting energy,
    calculated as a function of the dot minor diameter $l_y$, with
 fixed dot major diameter $l_x=20$~nm and dot height $l_z=3$~nm for (a)
    GaAs QD,
    (b) InAs QD; and with $l_x=12$~nm and $l_z=2$~nm for (c) GaN QD.} 
  \label{figtw13}
\end{figure}

In order to investigate the relative importance of the heavy-hole, light-hole
and split-off bands to the e-h exchange interaction, the doublet splitting
energy and the BD exchange splitting are calculated by
taking into account: (i) the heavy-hole band 
only; (ii) heavy-hole and light-hole bands; and (iii) all the 
three valence bands, separately. 
In Fig.~\ref{figtw13}, we plot the doublet splittings
originated from the three valence bands separately. The contribution of the
heavy-hole band is calculated by including the terms of e-h exchange
interaction derived from the heavy-hole band only as in Case (i). 
The contribution from the
light-hole (split-off) band is obtained by subtracting the splitting calculated
from Case (i) [Case (ii)] from the one from Case (ii) [Case (iii)]. 
It is noted that the doublet splitting energy is decreased by further including the
  light-hole and split-off bands. Therefore
in Fig.~\ref{figtw13}, the contributions from the light-hole and split-off bands are multiplied by
  $-1$.  Moreover, for GaN QDs, the structure splittings 
are calculated by setting $\Delta E_{\rm
  SR}=20~\mu$eV.\cite{GaN-sr}
As shown in the figure, the
contribution of the heavy-hole band is much 
larger than those from the other two bands. The doublet splitting energy is
slightly changed by further including the light-hole and split-off bands as in
Case (ii) and (iii), with the split-off band contributing
 most of the change. Under all
size parameters adopted, the doublet splitting energy is only
changed by less than 2~$\%$ in both GaAs and GaN QDs, and less than $7.2~\%$ in
InAs QDs. One can hence conclude that the terms derived from the heavy-hole
band dominate the off-diagonal matrix elements of the LR exchange
interaction [Eq.~(\ref{equ:Ham_long23}) and its Hermitian conjugate],
i.e.,  $|S_{ij}|\ll 1$.
And they are also much larger than the off-diagonal matrix elements of the SR
exchange interaction [i.e., Eq.~(\ref{equ:Ham_short23}) and its Hermitian 
conjugate].

Similarly, we find that the contribution of light-hole
and split-off bands to the BD exchange splitting is also small, in the order of
$1~\mu$eV, which is much 
smaller than  100s of $\mu$eV from the 
heavy-hole band for all three materials
under investigation. This can be understood
in the same way: according to  
Eqs.~(\ref{equ:Ham_long})-(\ref{equ:Ham_short33}), for the LR exchange
interaction, terms without $S_{ij}$ and $S^{\prime \ast}_{ij}$ should also dominate 
the diagonal matrix elements of the LR exchange 
interaction which contribute to the BD exchange
splitting. Moreover, for the SR exchange interaction,  the
diagonal matrix elements [Eqs.~(\ref{equ:Ham_short22}) and 
(\ref{equ:Ham_short33})]  are not affected by the inclusion
of the light-hole and split-off bands up to the order under consideration. 

In short, both the light-hole and split-off bands 
are negligible when the band-edge exciton fine structure is investigated 
in cubic III-V semiconductor QDs with strong confinement along 
the [001] direction. This further demonstrates the feasibility of 
treating the light-hole and split-off bands
 perturbatively through the L\"owdin partition. 
However, as discussed after Eqs.~(\ref{equ:V1}) and
(\ref{equ:V2}), the confinement-induced valence band mixing is crucial in
understanding some exciton properties, e.g., the observability of the dark
exciton\cite{Bayer1,Bayer3,Gorycal,Nirmal} and the degree of the
 linear polarization of  QD emission.\cite{Leger}


\section{Exciton spin relaxation in GaN QDs}
In this section, we study the exciton spin relaxation in single GaN QDs. The
relaxation rate is calculated from the Fermi golden rule with the exciton
eigenfunctions and eigenenergies obtained from the exact 
diagonalization method.\cite{note7} Since the ground 
state of the bright exciton is polarized along the major axis of the potential
ellipse, the relaxation rate is calculated from the upper
state of the doublet to the lower one, i.e., $|Y\rangle\rightarrow |X\rangle$ for $l_x > l_y$ and
$|X\rangle\rightarrow |Y\rangle$ for $l_x<l_y$. 

\begin{figure}[htb]
  \begin{center}
    \includegraphics[width=7.5cm]{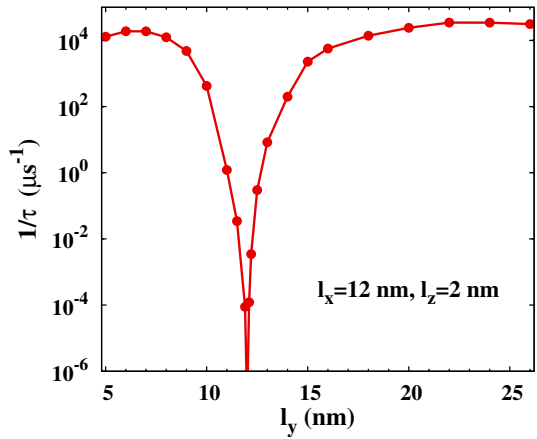}
  \end{center}
  \caption{The dependence of relaxation rate between the lowest
    $|X\rangle$ and $|Y\rangle$ bright exciton states on the dot major/minor
    diameter $l_y$ with fixed minor/major diameter $l_x=12$~nm and  dot height $l_z=2$~nm.}
  \label{figtw14}
\end{figure}

In the calculation, we fix the major/minor diameter $l_x=12$~nm and the dot
height $l_z=2$~nm.\cite{Marie} The shape dependence of the exciton
spin relaxation rate is studied by varying the minor/major
diameter $l_y$ and the results are plotted in
Fig.~\ref{figtw14}. One observes from the figure that the relaxation rate
between the lowest $|X\rangle$ and $|Y\rangle$ 
exciton states shows strong dependence on the dot anisotropy. In the regime
of large anisotropy, the relaxation rate reaches $10^{4}~\mu$s$^{-1}$, which in turn
gives the exciton spin relaxation time in the order of 10~ps. In the vicinity of
$l_y=l_x$, the relaxation rate
decreases quickly with decreasing dot anisotropy. When the QD
approaches the circular shape, the rate of change is even larger. The
relaxation rate decreases down to less than $100$~s$^{-1}$ and tends to zero when
$l_y$ approaches $l_x$. This calculated long exciton spin
relaxation time, especially in the range of $l_y$ 11$\sim$13~nm, is supported by 
the latest experiment.\cite{Marie} 

It is pointed out that the behavior of spin relaxation rate
in the range of $l_y=7\sim$16~nm mostly results from the anisotropy dependence of
doublet splitting energy shown in Fig.~\ref{figtw8}(b). 
From Eq.~(\ref{equ:relax}),  one has
\begin{equation}
  \Gamma_{i\to f}=\frac{2\pi}{\hbar}\sum_{\lambda}\int d\theta d\phi
  q_{\lambda}^2|M_{{\bf q}_{\lambda}}|^2|\langle f|\chi({\bf q}_{\lambda})|i\rangle|^2.
\end{equation}
When the wave vector $q_{\lambda}$ is not too large, the value of the relaxation
rate is mainly modulated by the factor $q_{\lambda}^2|M_{{\bf
    q}\lambda}|^2$, which increases with increasing $q_{\lambda}$ for all three
channels under consideration.  
The wave vector of the acoustic phonon is given by
$q_{\lambda}=\frac{\Delta E}{\hbar v_{\lambda}}$, with $v_{\lambda}$ representing
the sound velocity and $\Delta E$ denoting the 
phonon energy, which is equal to the difference 
of eigenenergies between the two bright exciton states 
considered, i.e., the doublet splitting energy. One observes from
Fig.~\ref{figtw8}(b) that qualitatively, the doublet splitting energy is
proportional to the dot anisotropy. So large dot anisotropy indicates large
phonon wave vector, which in turn results in  
the anisotropy dependence of relaxation rate as shown in
Fig.~\ref{figtw14} in the range of $l_y=7\sim 16$~nm. 

With further decrease of $l_y$ down to less than 7~nm, the exciton spin relaxation 
rate reaches a maximum around $l_y=6$~nm, where the doublet splitting energy 
is about $730~\mu$eV which corresponds to the situation that the wavelength of the emissive phonon
becomes comparable with twice of the lateral dot size.\cite{Ka,Bockelmann} 
On the other side, when $l_y$ increases over 16~nm, the increase of the
relaxation rate becomes slower and when $l_y$ reaches 24~nm, the relaxation
rate also reaches a maximum. The underlying physics is different from the
previous one. Here, the maximum is attributed to the interplay of the dot
anisotropy and the strength of the lateral confinement. On one hand, the increase of $l_y$ gives a
more anisotropic dot shape which tends to increase the doublet splitting
energy;
on the other hand, the confinement along the
$y$-direction becomes weaker with larger $l_y$, which tends to decrease the overlap of the
electron and hole wave functions and reduce the doublet splitting energy
according to Eq.~(\ref{equ:Ham_long23}). The competing leads to the maximum.

\begin{figure}[htb]
  \begin{center}
    \includegraphics[width=7.cm]{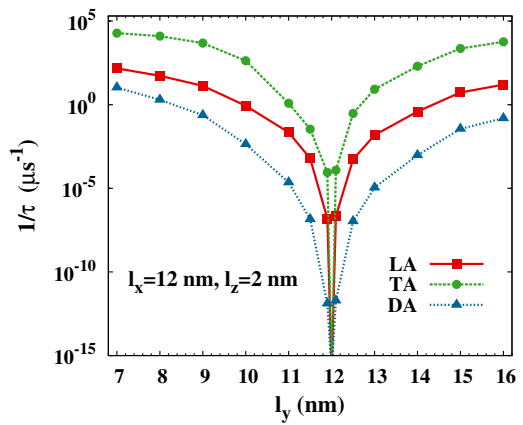}
  \end{center}
  \caption{(Color online) The relaxation rate limited by the
electron/hole-longitudinal acoustic phonon scatterings due to the deformation
potential~(DA) and the piezoelectric coupling ~(LA),  and by the 
electron/hole-transverse acoustic phonon scattering due to the piezoelectric
coupling~(TA). The relaxation rates are plotted as function of the dot major/minor
diameter $l_y$ with fixed minor/major diameter $l_x=12$~nm and dot height $l_z=2$~nm.} 
  \label{figtw15}
\end{figure}

In addition, the relative importance of the three channels contributing to the 
exciton relaxation rate is investigated. We take into account one channel at a time
and the results are plotted in Fig.~\ref{figtw15}. It is seen 
that the relaxation rate limited by the electron/hole-longitudinal acoustic phonon
scattering due to the piezoelectric field 
coupling is always 1 to 3 orders of magnitude larger than that due to the
deformation potential, but is 2 to 
3 orders of magnitude smaller than the one limited by the electron/hole-transverse
acoustic phonon scattering due to the piezoelectric coupling. So
the transverse acoustic phonon-emission process dominates the exciton spin
relaxation between the lowest $|X\rangle$ and $|Y\rangle$ bright exciton
states in GaN QDs. 
 
\section{SUMMARY}
In summary, we have established a general scheme to investigate the exciton fine
structure and spin relaxation in cubic III-V semiconductor QDs. 
A $12\times 12$ matrix representation
of the exciton Hamiltonian corresponding to the LR and SR exchange interactions
is derived by taking into account the conduction band 
$\Gamma_6^c$, the heavy-hole and light-hole 
bands $\Gamma_8^v$ and the split-off band $\Gamma_7^v$, where the 
split-off band has never been
included explicitly to investigate the exciton properties. 
In the case with strong
confinement in one direction (the [001] direction in 
this paper), the L\"owdin partitioning method is employed to 
take account of the
confinement-induced band mixing and a four-band
Hamiltonian of the e-h exchange interaction is derived. We use these formulae in
the study of the relative importance of the heavy-hole, light-hole and split-off
bands to the exciton fine structure.  
We find that the contribution of the split-off band is a little larger than
that of the light-hole band, but both are {\em negligible} when considering
the exciton fine structure in GaAs, InAs and GaN QDs. This behavior in GaN QDs
is unexpected since a significant effect of the 
split-off band is expected due to the
large band gap and the small spin-orbit splitting in cubic GaN. We attribute
this to the confinement-induced subband splitting due to the different effective masses
of the heavy, light and split-off holes in the $z$ direction. 

In our approach, the direct Coulomb interaction is treated 
unperturbatively. We find that
the strength of the direct Coulomb interaction strongly affects the doublet
splitting energy (hence also the BD exchange splitting). 
We also show that previous works in which the direct Coulomb interaction
was treated perturbatively vastly underestimate the doublet splitting.
We demonstrate  that the exact inclusion
of the direct Coulomb interaction is important for excitons in the weak
confinement regime.\cite{note2}

We also discuss the size and shape dependences 
of the doublet splitting energy
and the BD exchange splitting. Strong anisotropy 
dependence of the doublet splitting
energy is reported, which agrees with the former theoretical and experimental
works on GaAs and InAs QDs.\cite{Seguin,Taka2} The size dependences of the
doublet splitting energy and the BD exchange 
splitting are investigated by performing the size-scaling analysis.
The behavior of the variation of the fine structure splittings 
with the dot size is well explained by the scaling rules established. 
The doublet splitting
  energy in cubic GaN QDs increases with the increase of dot anisotropy
and/or the decrease of dot size, varying from 0 to 100s of
$\mu$eV, and reaching up to several meV 
for extremely small dot size and large dot
anisotropy. Our results are well supported by recent
experimental findings\cite{Marie,Kindel} but call for
more experimental works. 
The still undetermined singlet-triplet splitting $\Delta E_{\rm SR}$ 
in cubic GaN can be fit from future experimental data on the BD exchange
splitting, but the uncertainty does not affect the conclusions of this
paper.   

We  investigate the relative importance of 
the LR and SR exchange interactions to the exciton fine
structure. The LR exchange interaction is identified as the 
origin of the exciton
doublet structure, which is in agreement with that by 
Takagahara\cite{Taka2} where only
the heavy- and light-hole bands were included. We show that the LR
exchange interaction, which is 
 absent in many previous works,\cite{Efros,Horo}
 contributes to the splitting between the 
bright and dark exciton states, even 
in circular QDs where the doublet splitting vanishes. 
The contribution of the LR exchange interaction to the BD exchange splitting
is smaller than that of the SR one in GaAs QDs but is
comparable in InAs and GaN QDs. Our
calculations also demonstrate that the
relative importance of the LR and SR exchange interactions to the BD exchange
splitting can exchange with the variation of  the dot lateral size.

The exciton spin relaxation in cubic GaN QDs is also
investigated. We find that the
exciton spin relaxation rate strongly depends on the dot anisotropy (relaxation
rate from $10^4~\mu$s$^{-1}$ down to less than $10^{-4}~\mu$s$^{-1}$ is reported).
 In the small anisotropy regime, the long exciton spin relaxation time obtained
(longer than 100~ns) is in good agreement with recent experiment
by Lagarde {\em et al.}.\cite{Marie} The
electron/hole-transverse acoustic phonon scattering due the piezoelectric field
is recognized as the dominant magnetism of the exciton spin relaxation.   

Finally, we address the possible extensions of our work: (a) Other than the
shape anisotropy of QDs, strain anisotropy can also result in the doublet
splitting, which is not included in the present investigation. (b)
Within our model, the infinite square well potential is employed as the
confinement along the $z$ direction and only the lowest electron and hole subband is
included. As pointed out in Ref.~\onlinecite{note6}, in the case of extremely
small QDs, one needs to switch to other model potential or approach such as the 
psudopotential approximation\cite{Zunger2,Bester1,Bester2} to obtain more accurate results.
Meanwhile, for QDs, the vertical height of which is not much smaller than its lateral size or the exciton Bohr
radius, the multi-subband effect has to be included. (c) The fine
structure splittings may also be modulated by external electric and/or magnetic
fields which are not discussed in this paper. (d) Our model can also be
extended to investigate the initial optical spin polarizations.\cite{Marie2,Pfalz,Kowa}

\begin{acknowledgments}
This work was supported by the Natural Science Foundation of China
under Grant No.~10725417. We would like to thank X. Marie and I. E. 
Ivchenko for stimulating discussions. One of the authors (H.T.) 
would like to thank K. Shen for valuable discussions.
\end{acknowledgments}

\begin{appendix}

\section{The Exciton Hamiltonian}

Here we write down the explicit form of the exciton 
Hamiltonian $H^{\rm eh}_{m^{\prime}n^{\prime}
  \atop mn}\left({\bf r}^{\prime}_1\ {\bf r}^{\prime}_2 
\atop {\bf r}_1\ {\bf r}_2\right)$ in
Eq.~(\ref{equ:SE1}), which is recovered following the way laid out by 
    Pikus and Bir\cite{Bir1,Bir2}
\begin{eqnarray}
  \nonumber
  && H^{\rm eh}_{m^{\prime}n^{\prime} \atop mn}\left({\bf r}^{\prime}_1\ {\bf r}^{\prime}_2 \atop {\bf r}_1\ {\bf
        r}_2\right) = \left[  H^e_{m^{\prime}m}({\bf k}_1)\delta_{n^{\prime}n}+H^h_{n^{\prime}n}({\bf
    k}_2)\delta_{m^{\prime}m} \right. \\
  \nonumber
  && \ \ \ \ \ \ \left. \mbox{}+U^{\rm eh}({\bf r}_1-{\bf r}_2)\delta_{m^{\prime}m}\delta_{n^{\prime}n}\right]\delta({\bf
        r}_1-{\bf r}^{\prime}_1)\delta({\bf r}_2-{\bf r}^{\prime}_2)\\
  && \ \ \ \ \ \  \mbox{}+\Delta U^{{\rm ex}}_{m^{\prime}n^{\prime} \atop mn}
      \left({\bf r}^{\prime}_1\ {\bf r}^{\prime}_2 \atop {\bf r}_1\ {\bf r}_2\right), 
\label{equ:Ham_total}
\end{eqnarray}
where ${\bf k}=-i\bigtriangledown$ and
\begin{eqnarray}
 && \hspace{-0.3cm}U^{\rm eh}({\bf r}_1-{\bf r}_2)=-\frac{e^2}{4\pi
    \varepsilon_0\kappa |{\bf r}_1-{\bf r}_2|}, \label{equ:coulomb}\\
  \nonumber
  &&\hspace{-0.3cm} H^e_{m^{\prime}m}({\bf k}_1)= \Big[ E_m({\bf k}_0) +\frac{\hbar^2}{2m}{\bf k}_1^2\Big] +\frac{\hbar}{m}
  ({\bf k}_1\cdot{\bm \pi}_{m^{\prime}m}) \\ && \mbox{}+\frac{\hbar^2}{m^2}\sum_{n^{\prime\prime}}\frac{({\bf k}_1\cdot{\bm \pi}_{m^{\prime}n^{\prime\prime}})({\bf
      k}_1\cdot{\bm \pi}_{n^{\prime\prime}m})}{E^0_m-E^0_{n^{\prime\prime}}}, \label{equ:Ham_con}\\
  && \hspace{-0.3cm}H^h_{n^{\prime}n}({\bf k}_2)=-H^e_{\Theta n \Theta n^{\prime}}(-{\bf k}_2)=-H^e_{n^{\prime}n}({\bf
    k}_2). \label{equ:Ham_val}
  \end{eqnarray}
Here $H^e_{m^{\prime}m}$ is in the usual form derived from the ${\bf k}\cdot{\bf p}$
method up to the second order. We have ${\bm \pi} = {\bf p} +
\frac{\hbar}{4mc^2}\left[ {\bm \sigma} \times (\bigtriangledown V_0)\right]$
with $V_0$ standing for the lattice potential. ${\bm \pi}_{mn}$ denotes the matrix element of 
${\bm \pi}$ between the two Bloch functions $\psi_{m{\bf
      k}_0}({\bf r})$ and $\psi_{n{\bf
      k}_0}({\bf r})$. In this paper, lattice potential $V_0$ is assumed to
  have spherical symmetry. $\Theta$ stands for the time reversal
operator.

The e-h exchange interaction is decomposed into LR and SR parts:
\begin{equation}
  \Delta U^{{\rm ex}}_{m^{\prime}n^{\prime} \atop mn}\left({\bf r}^{\prime}_1\ {\bf r}^{\prime}_2 \atop {\bf r}_1\ {\bf
      r}_2\right) =H^{{\rm LR}}_{m^{\prime}n^{\prime} \atop mn}\left({\bf r}^{\prime}_1\ {\bf r}^{\prime}_2 \atop {\bf r}_1\ {\bf
      r}_2\right)+H^{{\rm SR}}_{m^{\prime}n^{\prime} \atop mn}\left({\bf r}^{\prime}_1\ {\bf r}^{\prime}_2 \atop {\bf r}_1\ {\bf
      r}_2\right),\label{equ:Ham_exchange_all}
\end{equation}
with
\begin{eqnarray}
\nonumber
&&  \hspace{-0.8cm}H^{{\rm LR}}_{m^{\prime}n^{\prime} \atop mn}\left({\bf r}^{\prime}_1\ {\bf r}^{\prime}_2 \atop {\bf r}_1\ {\bf
      r}_2\right)=-\sum_{\alpha\beta}Q^{\alpha\beta}_{m^{\prime}\Theta n \atop
   \Theta n^{\prime}m}\frac{\partial^2}{\partial{\bf r}^\alpha_1\partial{\bf r}^\beta_1}
  U({\bf r}_1-{\bf r}^{\prime}_2) \\ &&  \ \ \ \ \ \ \mbox{} \times \delta({\bf r}_1-{\bf r}_2)\delta({\bf r}^{\prime}_1-{\bf
    r}^{\prime}_2), \label{equ:Ham_exchange_long} \\
\nonumber
  && \hspace{-0.8cm}H^{{\rm SR}}_{m^{\prime}n^{\prime} \atop mn}\left({\bf r}^{\prime}_1\ {\bf r}^{\prime}_2 \atop {\bf r}_1\ {\bf
      r}_2\right)={\cal V}U_{m^{\prime}\Theta n \atop \Theta n^{\prime}m}\delta({\bf r}_1-{\bf r}_2)\delta({\bf
    r}_1-{\bf r}^{\prime}_1) \\ && \ \ \ \ \ \ \mbox{}\times \delta({\bf r}_2-{\bf r}^{\prime}_2).\label{equ:Ham_exchange_short}
\end{eqnarray}
In these equations, we have
\begin{eqnarray}
  \hspace{-0.8cm}  Q^{\alpha\beta}_{m^{\prime}\Theta n \atop \Theta
    n^{\prime}m}&=&\frac{\hbar^2}{m^2}\frac{\pi^\alpha_{m^{\prime}\Theta n^{\prime}}\pi^\beta_{\Theta nm}}{(E_m^0-E_n^0)(E_{m^{\prime}}^0-E_{n^{\prime}}^0)},
  \label{equ:Q_2}\\
U({\bf r})&=&\frac{e^2}{4\pi
    \varepsilon_0\kappa |{\bf r}|},
\end{eqnarray} 
and $E_s^0$ represents the eigenenergy of the $s$ 
band at the point ${\bf k}={\bf k}_0$; ${\cal V}$ is the volume of the bulk material which
comes from the normalization of the Bloch function and
\begin{eqnarray}
\nonumber
  &&U_{m^{\prime}\Theta n \atop \Theta n^{\prime}m} = \frac{1}{{\cal V}^2}\int\int \psi^\ast_{m^{\prime}{\bf k}_0}({\bf
    r}_1)(\Theta\psi_{n{\bf k}_0}({\bf r}_2))^\ast U({\bf r}_1-{\bf r}_2) \\ &&
  \ \ \ \ \ \ \mbox{}\times\Theta\psi_{n^{\prime}{\bf k}_0}({\bf
    r}_1)\psi_{m{\bf k}_0}({\bf r}_2)d{\bf r}_1d{\bf r}_2.
\label{equ:Def_U}
\end{eqnarray}

\section{The Bloch Functions}
The Bloch functions of the $\Gamma_6^c$ conduction band take the form\cite{Winkler}
\begin{equation}
  |\frac{1}{2},\frac{1}{2}\rangle_c = |S\rangle\alpha, \ 
  |\frac{1}{2},-\frac{1}{2}\rangle_c = |S\rangle\beta. 
\label{equ:Wave_c1}
\end{equation}
where $\alpha$ ($\beta$) denotes the spin-up (down) state 
and $|S\rangle$ represents
  the $s$-like conduction band Bloch function. The Bloch functions of
the $\Gamma_8^v$ and $\Gamma_7^v$ valence bands 
are written as\cite{Winkler}
\begin{eqnarray}
&&|\frac{3}{2},\frac{3}{2}\rangle_v =
    -\frac{1}{\sqrt{2}}(|X\rangle+i|Y\rangle)\alpha,\\
&&|\frac{3}{2},\frac{1}{2}\rangle_v =
\frac{1}{\sqrt{6}}\big[-(|X\rangle+i|Y\rangle)\beta+2|Z\rangle\alpha\big]
,\\
&&|\frac{3}{2},-\frac{1}{2}\rangle_v = \frac{1}{\sqrt{6}}\big[(|X\rangle-i|Y\rangle)\alpha+2|Z\rangle\beta\big],\\
  &&|\frac{3}{2},-\frac{3}{2}\rangle_v = \frac{1}{\sqrt{2}}(|X\rangle-i|Y\rangle)\beta,\\
  &&|\frac{1}{2},\frac{1}{2}\rangle_v = -\frac{1}{\sqrt{3}}\big[({|X\rangle+i|Y\rangle})\beta+|Z\rangle\alpha\big],\\
  &&|\frac{1}{2},-\frac{1}{2}\rangle_v = -\frac{1}{\sqrt{3}}\big[({|X\rangle-i|Y\rangle})\alpha-|Z\rangle\beta\big],
\end{eqnarray}
where $|X\rangle$, $|Y\rangle$ and $|Z\rangle$ are the $p$-like valence band
  Bloch functions which are real according to the phase convention in accordance
  with the time-reversal symmetry. 
After taking the time reversal operation, we have
\begin{eqnarray}
   && \hspace{-0.8cm} \Theta|\frac{3}{2},\frac{3}{2}\rangle_v  = -\frac{1}{\sqrt{2}}(|X\rangle-i|Y\rangle)\beta,\\
  && \hspace{-0.8cm} \Theta|\frac{3}{2},\frac{1}{2}\rangle_v  = \frac{1}{\sqrt{6}}\big[(|X\rangle-i|Y\rangle)\alpha+2|Z\rangle\beta\big],\\
  && \hspace{-0.8cm} \Theta|\frac{3}{2},-\frac{1}{2}\rangle_v = \frac{1}{\sqrt{6}}\big[(|X\rangle+i|Y\rangle)\beta-{2}|Z\rangle\alpha\big],\\
  && \hspace{-0.8cm} \Theta|\frac{3}{2},-\frac{3}{2}\rangle_v  = -\frac{1}{\sqrt{2}}(|X\rangle+i|Y\rangle)\alpha,\\
  && \hspace{-0.8cm} \Theta|\frac{1}{2},\frac{1}{2}\rangle_v =
    \frac{1}{\sqrt{3}}\big[({|X\rangle-i|Y\rangle})\alpha-|Z\rangle\beta\big],\\
  && \hspace{-0.8cm} \Theta|\frac{1}{2},-\frac{1}{2}\rangle_v = -\frac{1}{\sqrt{3}}\big[({|X\rangle+i|Y\rangle})\beta+|Z\rangle\alpha\big].
\end{eqnarray}

\section{Construction of basis functions with direct Coulomb interaction
  explicitly included}

With the confinement, the diagonal part of the exciton Hamiltonian $H^{\rm
    eh}_{m^{\prime}n^{\prime} \atop mn}\left({\bf r}^{\prime}_1\ {\bf r}^{\prime}_2 \atop {\bf r}_1\ {\bf
      r}_2\right)$, i.e., excluding the e-h exchange interaction, can be
written into
\begin{equation}
    H_D=H_e+H_h+H_{\rm Coulomb}+H_{\rm confinement},\label{equ:HD}
\end{equation}
where $H_e$ is the electron Hamiltonian in the form $H_e=\hbar^2k_e^2/2m_e^\ast$
with $m_e^\ast$ being the effective mass of the conduction electron and $H_h$ is
the hole Hamiltonian. From Eq.~(\ref{equ:Ham_hole}), we see that $H_h$ is 
diagonal in the 4$\times$4 matrix representation and the quasi-spins are not
coupled. So in the following we omit the spin degree of freedom of both 
electron and hole. $H_{\rm Coulomb}$ and $H_{\rm confinement}$ are the direct Coulomb
interaction and the confinement potential, given in Eq.~(\ref{equ:coulomb}) and 
Eqs.~(\ref{equ:con1})-(\ref{equ:con2}). The eigen equation for the envelope basis function is constructed as
\begin{eqnarray}
  H_D | {\rm eh} \rangle &=& E | {\rm eh} \rangle, \label{equ:eq1}\\
  \langle {\bf r}_1,{\bf r}_2 | {\rm eh} \rangle &=& f ({\bf r}_1,{\bf r}_2 ).
\end{eqnarray}
When a strong confinement is applied along the 
$z$-direction so that only the lowest electron/hole
subband is relevant, one has
\begin{equation}
   f ({\bf r}_1,{\bf r}_2 )=\Phi ({\bf r}_{1||},{\bf r}_{2||} )\xi(z_1)\zeta(z_2),
\end{equation}
where $\xi(z) [\zeta(z)]=\sqrt{\frac{2}{l_z}}{\rm sin}(\frac{\pi z}{l_z})$ 
stands for the electron (hole) envelope function in the $z$-direction. After
multiplying both sides of Eq.~(\ref{equ:eq1}) 
with $\xi(z_1)\zeta(z_2)$ and integrating over $z_1$ and $z_2$, one comes to
\begin{equation}
\tilde{H}_D\Phi ({\bf r}_{1\parallel},{\bf r}_{2\parallel} ) =
E\Phi ({\bf r}_{1\parallel},{\bf r}_{2\parallel}),
\end{equation}
with
\begin{eqnarray}
  \nonumber
 &&\hspace{-0.7cm}\tilde{H}_D=\frac{p_{1\parallel}^2}{2m_e^\ast}+\frac{p_{2\parallel}^2}{2m_{h\parallel}^\ast}+\frac{1}{2}m_e^\ast(\omega_{xe}^2x_1^2+\omega_{ye}^2y_1^2)\\
 &&
  \ \ \ \hspace{-0.6cm} \mbox{}+
  \frac{1}{2}m_{h\parallel}^\ast(\omega_{xh}^2x_2^2+\omega_{yh}^2y_2^2)+V({\bf r}_{\parallel})+E_z^e+E_z^h,
  \end{eqnarray}
in which
\begin{eqnarray}
  &&\hspace{-0.9cm}m_{h\parallel}^\ast=\frac{m_0}{\gamma_1+\gamma_2},\ \ 
  E_z^e=\frac{\hbar^2\pi^2}{2m_e^\ast l_z^2},\  
  E_z^h=\frac{\hbar^2\pi^2}{2m_{h,z}^\ast l_z^2}, \\ 
 &&\hspace{-0.9cm}m_{h,z}^\ast=\frac{m_0}{\gamma_1-2\gamma_2}, \ \ {\bf
    r}_{\parallel}=(x_1-x_2,y_1-y_2), 
  \\
\nonumber
  &&\hspace{-0.9cm} V({\bf
    r}_{\parallel})=\frac{4}{l_z^2}\int_0^{l_z}dz_1\int_0^{l_z}dz_2U^{\rm eh}({\bf r}_1-{\bf
    r}_2)\\ &&\hspace{0.2cm}  \mbox{}\times[{\rm sin}({\pi
  z_1}/{l_z}){\rm sin}({\pi z_2}/{l_z})]^2.
  \end{eqnarray}
Here, $m_{h\parallel}^\ast$ and $m_{h,z}^\ast$ are
 the effective masses of the heavy-hole
in the plane and in the $z$-direction, respectively;
 $E_z^e$ and $E_z^h$ are the subband energies
resulting from the strong confinement in the $z$-direction.

After separating  the coordinates of electron-hole pair in the plane into
the center-of-mass and relative parts: 
\begin{equation}
  \begin{array}{ll}
    {\bf r}_{\parallel}={\bf r}_{1\parallel}-{\bf r}_{2\parallel} \\
    {\bf R}_{\parallel}=\frac{m_e^\ast{\bf r}_{1\parallel}
+m_{h\parallel}^\ast{\bf r}_{2\parallel}}{m_e^\ast+m_{h\parallel}^\ast}
    \end{array} \Longrightarrow\left\{
    \begin{array}{ll}
    {\bf K}_{\parallel}={\bf k}_{1\parallel}+{\bf k}_{2\parallel} \\
    {\bf k}_{\parallel}=\frac{m_{h\parallel}^\ast{\bf k}_{1\parallel}-
m_e^\ast{\bf k}_{2\parallel}}{m_e^\ast+m_{h\parallel}^\ast}
    \end{array},
    \right. 
 \end{equation}
 $\tilde{H}_D$ is separated into two parts: 
$\tilde{H}_D=\tilde{H}_0+\tilde{H}^{\prime}$, where 
\begin{eqnarray}
\nonumber
    &&\hspace{-0.3cm}\tilde{H}_0=\frac{\hbar^2}{2m_{\mu}}(k_x^2+k_y^2)+
    \frac{1}{2}m_{\mu}\omega_{\rm ave}^2(x^2+y^2)\\
\nonumber
&& \mbox{}+\frac{1}{2}{m_0^2}\big(\frac{1}{m_e^\ast}+\frac{1}{m_{h\parallel}^\ast}\big)(\omega_{x0}^2X^2
      +\omega_{y0}^2Y^2)\\&& \mbox{}+\frac{\hbar^2}{2(m_e^\ast+m_{h\parallel}^\ast)}
      (K_x^2+K_y^2)+V(r_{\parallel}), \label{equ:Ham_0}
\end{eqnarray}
in which 
  \begin{eqnarray}    
   &&\hspace{-1cm}m_{\mu}=\frac{m_e^\ast m_{h\parallel}^\ast}{m_e^\ast+m_{h\parallel}^\ast},
   \ \ \omega_{\rm ave}^2=\frac{\omega_{xr}^2+\omega_{yr}^2}{2},\\ 
  &&\hspace{0cm}\omega_{xr}=\hbar \Big({\frac{m_e^\ast m_{h\parallel}^\ast}
{\sqrt{m_e^{\ast 2}-m_e^{\ast}m_{h\parallel}^\ast+m_h^{\ast 2}}}l_x^2}\Big)^{-1},\\    
    &&\hspace{0cm}\omega_{yr}=\hbar\Big({\frac{m_e^\ast m_{h\parallel}^\ast}
{\sqrt{m_e^{\ast 2}-m_e^{\ast}m_{h\parallel}^\ast+m_h^{\ast 2}}}l_y^2}\Big)^{-1}.
  \end{eqnarray}
$\tilde{H}^{\prime}$ is further constituted of two parts:
\begin{equation}
\tilde{H}^{\prime}=H_{L1}+H_{L2}, \label{equ:H_per}
\end{equation}
 with
\begin{eqnarray}
      &&\hspace{-0.4cm}H_{L1}=m_0^2(\frac{b}{m_e^\ast}-\frac{a}{m_{h\parallel}^\ast})(\omega_{x0}^2xX+\omega_{y0}^2yY),
\label{equ:HL1}  
  \\ &&\hspace{-0.8cm}H_{L2}=\frac{1}{2}m_{\mu}\left[
      (\omega_{xr}^2-\omega_{\rm ave}^2)x^2+(\omega_{yr}^2
-\omega_{\rm ave}^2)y^2\right],\label{equ:HL2}  
\end{eqnarray}
where
\begin{eqnarray}
   && \hspace{-0.3cm} a=\frac{m_e^\ast}{m_e^\ast+m_{h\parallel}^\ast},\ \ \ b=\frac{m_{h\parallel}^\ast}{m_e^\ast+m_{h\parallel}^\ast}, \\
    && \hspace{-0cm}\omega_{x0}=\frac{\hbar}{m_0l_x^2}, \ \ \ 
    \ \ \ \omega_{y0}=\frac{\hbar}{m_0l_y^2}.
  \end{eqnarray}

\begin{figure}[htb]
  \begin{center}
    \includegraphics[width=7.5cm]{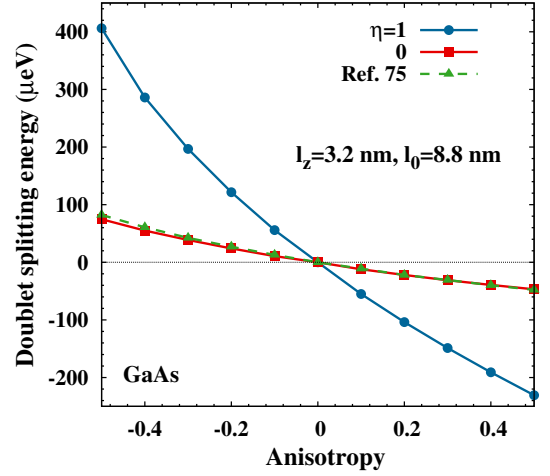}
  \end{center}
  \caption{(Color online) The dependence of doublet splitting energy on the dot
    anisotropy in an GaAs QD with $l_z=3.2$~nm and $l_0=8.8$~nm. $\eta$ is set
    at 1 or 0 to switch the direct Coulomb interaction on and off. The triangle
    points are taken from Fig.~2 in Ref.~\onlinecite{Kada}.}
  \label{figtw16}
\end{figure}

Treating $\tilde{H}^{\prime}$ perturbatively, we see from Eq.~(\ref{equ:Ham_0}) that the
center-of-mass and relative motions of the electron-hole pair are now decoupled
and we are able to write the in-plane wave function 
as $\Phi({\bf r}_{1\parallel},{\bf r}_{h\parallel}
)=\psi(X,Y)\varphi({\bf r}_{\parallel})$. Here $\psi$ is 
the eigenfunction of the 2D harmonic-oscillator potential
\begin{eqnarray}
\nonumber
&&\hspace{-0.5cm}\psi_{n_xn_y}(X,Y)=A_{n_x}A_{n_y}e^{-(\alpha_c^2X^2+\beta_c^2Y^2)/2}H_{n_x}(\alpha_cX)\\ 
&&\hspace{1.7cm}\mbox{}\times H_{n_y}(\beta_xY),
\end{eqnarray}
where
\begin{eqnarray}
&&\hspace{-0.3cm}\omega_{xc}=\frac{\hbar}{\sqrt{m_e^\ast m_{h\parallel}^\ast}l_x^2},
\ \   
\omega_{yc}=\frac{\hbar}{\sqrt{m_e^\ast m_{h\parallel}^\ast}l_y^2}, \label{equ:W-xy} \\
&&\hspace{-1cm}\alpha_c=\sqrt{\frac{(m_e^\ast+m_{h\parallel}^\ast)\omega_{xc}}{\hbar}},
\ \ 
\beta_c=\sqrt{\frac{(m_e^\ast+m_{h\parallel}^\ast)\omega_{yc}}{\hbar}},
\end{eqnarray}
with its eigenvalue being
$E_{n_xn_y}=(n_x+\frac{1}{2})\hbar\omega_{xc}+(n_y+\frac{1}{2})\hbar\omega_{yc}$. 
$A_{n_x}$ and $A_{n_y}$ are the normalization factors 
and $H_{n}(x)$ are the Hermit polynomials.

According to Eq.~(\ref{equ:Ham_0}), the relative part of the in-plane wave function can be expressed in polar
coordinates as $\varphi({\bf
  r}_{\parallel})_{mn}=\frac{1}{\sqrt{2\pi}}e^{im\phi}R_n(\rho)$ with $\rho=|{\bf
  r}_{\parallel}|$. $R_n(\rho)$ is obtained by numerically solving the radial
equation of the relative motion of the electron-hole pair in the real space.

Overall, the basis function for $F_{m^{\prime}n^{\prime}}({\bf r}^{\prime}_1,{\bf r}^{\prime}_2)$ in
Eq.~(\ref{equ:SE1}) is constructed as 
\begin{equation}
f_{mnn_xn_y} ({\bf r}_1,{\bf r}_2)=\psi_{n_xn_y}(X,Y)\varphi_{mn}({\bf
  r}_{\parallel})\xi(z_1)\zeta(z_2), \label{equ:basis} 
\end{equation}
 which is the eigenfunction of $H_0$ 
with
\begin{equation}
H_0=H_D-\tilde{H}^{\prime}.\label{equ:H0}
\end{equation}
 The exciton Hamiltonian $H^{eh}$
is diagonalized under the set
of basis \{$|f_{mnn_xn_y}\rangle\otimes|c_iV_j\rangle$\}, where $|c_iV_j\rangle$ acts as
quasi-spin. In this way, the exciton eigenstates and eigenvalues are obtained
with the direct Coulomb and the exchange interactions fully  accounted.

\section{Comparison with results in Ref.~69}
We calculate the doublet splitting energies in single GaAs QD with $\eta=0$ (the
direct Coulomb interaction is switched off) and with $\eta=1$ (the direct
Coulomb interaction is fully included). The length of dot major/ minor diameters 
are evaluated as $l_x=l_0/\sqrt{1+\xi}$ and $l_y=l_0\sqrt{1+\xi}$, which in turn 
give $\omega_{x\square}=\omega_{\square}^0(1+\xi)$ and $\omega_{y\square}
=\omega_{\square}^0/(1+\xi)$ with
$\omega_{\square}^0=\frac{\hbar}{m^\ast_\square l_0^2}$. Here $\square$ 
stands for electron or hole, $m^\ast_\square$ denotes the corresponding 
in-plane effective mass and $\xi=(l_y-l_x)/l_x$  
represents the dot anisotropy.
This is consistent with that in Ref.~\onlinecite{Kada}.
 The size parameters are chosen
carefully to simulate the model employed in Ref.~\onlinecite{Kada} according to
the characteristic energies induced by the confinement. In respect that the
confinement potentials are chosen separately for electron and hole in
Ref.~\onlinecite{Kada}, we deduce from the the parameters therein two sets of
size parameters. In the calculation, the size parameters are set between the
corresponding two values. We choose $l_z=3.2$~nm and 
$l_0=8.8$~nm.  The dependence of the doublet
splitting energy on the dot anisotropy for $\eta=0$ is shown in
Fig.~\ref{figtw16} (red curve with $\square$) which is
extremely close to that from Fig.~2 in Ref.~\onlinecite{Kada} (green
$\blacktriangle$). 
As we see, the absolute values of doublet splitting energy 
calculated with $\eta=1$ are
much larger than those with $\eta=0$.

\end{appendix}


\begin{thebibliography}{0}
\bibitem{note0} Here the heavy-, light- and split-off-hole exciton states
are defined in the approximation of decoupled heavy-hole, light-hole and
split-off bands, i.e., the band couplings are neglected or treated
perturbatively.  For the latter case, the heavy-hole, light-hole and
 split-off-hole states are denoted by their main components, which also
    include mixings of each other.
\bibitem{note1}During the procedure, special attention
should be paied to the following point: The Bloch functions are given in
different representaions due to the different phase conventions employed in the literature. The
convention used to derive the matrix representation of the e-h exchange
interaction must be the same to the one according to which the Luttinger
Hamiltonion is given, as long as we are going to employ both of them together.
\bibitem{note2}It is noted that even though the exciton Bohr radius in bulk InAs is
    larger than the dot size in our calculation which is in the strong 
    confinement regime according to the criterion commonly used in the
    literature,\cite{Efros,Ivchenko3} the
    exciton binding energy is found to be greatly enhanced
    from the bulk value (15~meV in the QD in constrast to 2.1~meV in bulk) due
    to the quantum confinement effect  
    and becomes comparable to the level spacing induced by the lateral
    confinement. This means that even the rough criterion indicates that
    nanostructures are in the strong confinement regime, the direct Coulomb
    interaction can still be too large to be treated perturbatively.
\bibitem{note3}For QDs investigated in this paper, the exciton binding energy
   is around 20$\sim$30~meV in GaAs QD, 10$\sim$20~meV in InAs QD and
   60$\sim$85~meV in GaN QD. The results in GaAs and InAs QDs are consistent
   with Ref.~60 and Ref.~96 accordingly.
\bibitem{note8}
It is noted that  the dark exciton doublet splitting is not
 described in our model. It can be taken into account by including the
 second order terms of $S_{ij}$ in Eqs.~(\ref{equ:Ham_long}) and
  (\ref{equ:Ham_ss}) or higher orders of $S_{ij}$ 
in $k_x$ and $k_y$.
\bibitem{note4} It is noted that in Ref.~\onlinecite{Taka2}, the
doublet splitting is calculated by choosing 
$l_x=20$~nm, $l_y=15$~nm and $l_z=3$~nm 
to fit the experiment data.\cite{Gammon} The small difference
of $l_y$ results
from the different forms of confinement potential.
\bibitem{note9}We note that other factors, e.g., the anisotropic strain, may
  also affect the doublet splitting energy.\cite{Bardoux,Kindel} But as  shown in our calculation,
 the dot geometry should play the key role in determining the dependence of the
doublet splitting on the exciton recombination energy.
\bibitem{note5}The level spacing between the ground and first excited dark
  exciton levels mainly comes from the excitation of the center-of-mass motion
  since the excitation energy of the relative motion is much higher. The exciton
  eigenstates and eigenenergies are obtained from the 
  excat diagonalization process and the original basis states are mixed with
  other levels. So the genuine scaling
  law of the level spacing can be a little different 
  from $\propto \frac{1}{L^2}$.
\bibitem{GaN-sr} The experimental value of singlet-triplet splitting $\Delta
  E_{\rm SR}$ in GaN is absent. Here we take it 20~$\mu$eV which is the
    singlet-triplet splitting value in GaAs 
temporarily, since they have close crystal structures.
\bibitem{note7}In the calculation, the singlet-triplet splitting $\Delta
  E_{\rm SR}$ is set at 20~$\mu$eV. We also calculate the exciton splin relaxation
  rate with $\Delta E_{\rm SR}=0~\mu$eV and the relaxation rate is found to be
  almost the same. This is understood  that the doublet splitting is
  dominated by the LR exchange interaction in GaN QDs and the SR exchange
  interaction is irrelevant.
\end{thebibliography}
\end{document}